\PassOptionsToPackage{table}{xcolor}
\documentclass[sigconf]{acmart}

\usepackage{colortbl}
\usepackage{array}

\usepackage{graphicx}
\usepackage{multirow}
\usepackage{subcaption}
\usepackage{booktabs,tabularx,array}

\usepackage[table]{xcolor} 
\newcolumntype{L}[1]{>{\raggedright\arraybackslash}p{#1}}
\newcolumntype{Y}{>{\raggedright\arraybackslash}X}

\AtBeginDocument{%
  }

\copyrightyear{2026}
\acmYear{2026}
\setcopyright{cc}
\setcctype{by}
\acmConference[CHI '26]{Proceedings of the 2026 CHI Conference on Human Factors in Computing Systems}{April 13--17, 2026}{Barcelona, Spain}
\acmBooktitle{Proceedings of the 2026 CHI Conference on Human Factors in Computing Systems (CHI '26), April 13--17, 2026, Barcelona, Spain}
\acmPrice{}
\acmDOI{10.1145/3772318.3790548}
\acmISBN{979-8-4007-2278-3/2026/04}
\begin{document}

\title{VR Calm Plus: Coupling a Squeezable Tangible Interaction with Immersive VR for Stress Regulation}

\author{He Zhang}
\orcid{0000-0002-8169-1653}
\affiliation{%
  \institution{The Pennsylvania State University}
  \city{University Park}
  \state{Pennsylvania}
  \country{USA}
  \postcode{16801}
}
\email{hpz5211@psu.edu}

\author{Xinyang Li}
\orcid{0000-0003-4081-604X}
\affiliation{%
  \institution{Tsinghua University}
  \city{Beijing}
  \country{China}
  \postcode{100084}
}
\email{lixinyang22@mails.tsinghua.edu.cn}

\author{Xingyu Zhou}
\orcid{0009-0002-9591-1191}
\affiliation{%
  \institution{City University of Hong Kong}
  \city{Hong Kong}
  \country{China}
}
\email{xyzhou62-c@my.cityu.edu.hk}

\author{Xinyi Fu}
\orcid{0000-0001-6927-0111}
\authornote{Corresponding author.}
\affiliation{%
  \institution{Tsinghua University}
  \city{Beijing}
  \country{China}
  \postcode{100084}}
\email{fuxy@tsinghua.edu.cn}

\renewcommand{\shortauthors}{Zhang et al.}

\begin{abstract}
While Virtual Reality (VR) is increasingly employed for stress management, most applications rely heavily on audio-visual stimuli and overlook the therapeutic potential of squeezing engagement. To address this gap, we introduce VR Calm Plus, a multimodal system that integrates a pressure-sensitive plush toy into an interactive VR environment. This interface allows users to dynamically modulate the virtual atmosphere through physical squeezing actions, fostering a deeper sense of embodied relaxation. We evaluated the system with 40 participants using PANAS-X surveys, subjective questionnaires, physiological measures (heart rate, skin conductance, pulse rate variability), and semi-structured interviews. Results demonstrate that, compared to a visual-only baseline, squeeze-based interaction significantly enhances positive affect and perceived relaxation. Physiological data further revealed a state of ``active relaxation'', characterized by greater reductions in heart rate and preserved autonomic flexibility (PRV), alongside sustained emotional engagement (GSR). Our findings highlight the value of coupling tangible input with immersive environments to support emotional well-being and offer design insights for future VR-based mental health tools.

\end{abstract}




\begin{teaserfigure}
\centering
\includegraphics[width=0.7\textwidth]{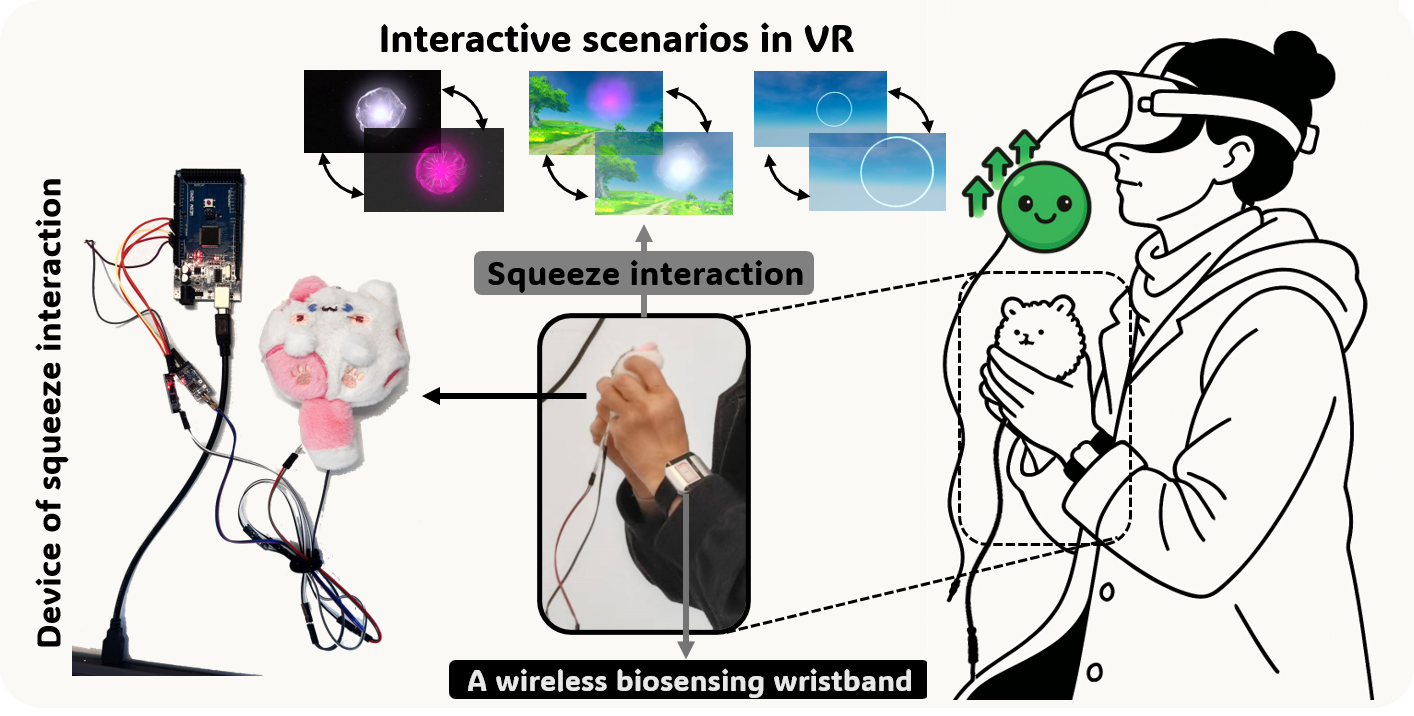}
\caption{VR Calm Plus system and interaction flow. On the left, a plush toy augmented with a pressure sensor and microcontroller serves as the squeeze-based input device. In the center, a close-up shows a participant squeezing the plush; a label at the bottom of this panel marks the wireless biosensing wristband that records physiological signals. At the top, example VR scenes illustrate how the virtual environment changes in response to squeezing. On the right, an illustration depicts a user wearing a VR headset while holding the plush and wristband, experiencing increased relaxation and positive affect.}
\Description{A soft plush toy instrumented with a pressure sensor and microcontroller forms the squeeze-based input device on the left. In the middle, a close-up photograph illustrates the squeeze interaction: as the participant compresses the plush toy, the sensor readings are streamed to the VR application. A label at the bottom of this close-up highlights the wireless biosensing wristband worn on the participant's wrist, which records physiological signals such as heart activity and skin conductance. At the top, example screenshots show interactive relaxation scenarios in VR, where visual elements dynamically respond to the intensity and rhythm of squeezing. On the right, an illustration depicts a user wearing a VR headset while holding the plush and wearing the wristband, accompanied by an icon representing increased positive affect.}
\label{fig:teaser}
\end{teaserfigure}
\maketitle

\section{Introduction}
Emotion, as a vital human characteristic, is closely linked to perception, attention, memory, and learning~\cite{Tyng2017}. The ability to regulate emotions plays an essential role in social life and affects everyone~\cite{Gross2002}. In today's fast-paced world and with the ever-increasing exchange of information, psychological issues such as stress and anxiety have become significant concerns affecting human well-being~\cite{oecd2018emotional,rahimnia2013emotional,huppert2009psychological}. As work, academic, social, and intimate relationship pressures, as well as the constant influx of stimuli from modern life—increase, individuals are more prone to experience anxiety or depression. When faced with negative emotions like stress, people typically seek practical outlets to vent and alleviate these feelings. However, without timely and effective emotion regulation strategies, such negative emotions can lead to psychological disorders and even trigger somatic reactions or adverse social events~\cite{CARL2013343}. This situation poses a serious challenge to both individual and societal welfare, underscoring the urgent need for effective stress management solutions.

In recent years, with the advancement of virtual reality (VR) technology, this powerful medium that blurs the boundaries between reality and virtual worlds~\cite{dudley2023inclusive} has gained popularity across multiple domains including entertainment~\cite{10535527}, education~\cite{9288462}, social relationship~\cite{zamanifard2023surprise}, and healthcare~\cite{alahmari2022outcomes}. VR provides users with immersive experiences~\cite{Skarbez2022Presence}, allowing them to obtain realistic sensations in constructed virtual environments (creating a sense of presence). Substantial research demonstrates that VR has significant effects on triggering and regulating emotions~\cite{9792298}, offering new opportunities to address the challenges mentioned above. Through immersive experiences that separate users from the real world, VR not only helps users engage in an independent virtual space but also promotes deeper levels of participation and interaction to achieve specific objectives. However, despite VR's advantages in immersion and presence, many existing VR applications still primarily rely on visual and auditory channels. While these can facilitate emotional regulation, they may significantly limit user engagement and experience. Therefore, exploring additional modalities in VR environments could become an important approach to further enhance user experience and optimize interaction design.

Beyond the common modalities in VR interaction (vision and hearing), touch, as one of the fundamental human perception methods, plays a vital role in emotional perception and expression. Against this background, this study aims to explore how squeeze feedback can be effectively integrated into VR systems to enhance user immersion and emotional regulation effects.

To this end, we designed VR Calm Plus (see in Figure~\ref{fig:teaser}), a squeezed-based interactive VR system intended to promote relaxation and stress relief by introducing positive squeeze experiences in VR interactions. The system employs a plush squeezable toy with built-in pressure sensors that trigger real-time dynamic changes in the virtual environment through the user's physical squeezing actions. This design not only enriches user interaction methods in virtual environments but also directly influences users' emotional states through intuitive squeezed-based feedback, helping to alleviate stress and promote relaxation. We conducted a detailed assessment of user experience feedback from 40 participants (22 females, 18 males) using a mixed-methods approach that combined pre- and post-experiment questionnaires (PANAS-X), physiological measures, and a well-designed subjective experience evaluation questionnaire, exploring the following key question:

\begin{itemize}
    \item \textbf{RQ.} How does the squeeze-based interaction system, VR Calm Plus, influence users' emotional states in a multimodal VR environment?
\end{itemize}

Our research makes four contributions to the field. First, we propose a novel VR system that combines tangible squeeze-based input with responsive visual feedback to enhance immersive experiences. Second, we provide new theoretical perspectives and empirical evidence for VR interaction design, especially for VR experiences incorporating elements of squeeze interactions, demonstrating that multisensory interaction can significantly alleviate negative emotions. Third, we present user-centered VR design recommendations for integrating physical and virtual interactions. Finally, this work helps advance the application and development of multisensory interaction technology in stress management and mental health fields, offering new insights and approaches for future research on VR-based stress management and mental health applications.

An earlier poster version of this work~\cite{11236289} presented high-level, summary results, including aggregated findings from pre- and post-PANAS-X measures and a user experience questionnaire. The present paper substantially extends this prior work with additional analyses and a more comprehensive discussion of design and theoretical implications.

\section{Related Work}

\subsection{Emotion Regulation, Positive Psychology and What is in VR}
Emotional regulation refers to how individuals influence which emotions they have, when they have them, and how they experience and express these emotions~\cite{gross1998emerging,gross2014emotion}. Building on this definition, researchers have expanded the concept, incorporating interpersonal relationships, internal physiological foundations, and external environmental factors as influences on emotional regulation~\cite{gross2015extended}. Gross~\cite{gross1998antecedent} categorizes emotional regulation processes based on their timing relative to emotion generation: antecedent-focused regulation and response-focused regulation. Among these, antecedent-focused strategies (particularly cognitive reappraisal) typically prove more effective than response-focused approaches. Cognitive reappraisal not only reduces the intensity of negative emotions but also enhances emotional adaptability by altering an individual's interpretation of situations, enabling effective responses to emotional challenges~\cite{bonanno2013regulatory}. In contrast, emotional suppression, while temporarily controlling external emotional expression, generally fails to modify internal emotional experiences and may lead to emotional accumulation and psychological problems~\cite{gross1993emotional,sheppes2015emotion,gross1998emerging}.

\subsubsection{Positive Psychology}
Over the past several decades, research and practice in emotion regulation have primarily focused on the mitigation and management of negative emotions, a direction that closely aligns with the traditional emphasis of psychopathology. However, with the rise of ``Positive Psychology''~\cite{seligman2000positive}, this research paradigm has undergone a significant shift. Positive psychology advocates for enhancing mental health and subjective well-being by cultivating, sustaining, and reinforcing positive emotional states, enabling individuals to better cope with everyday stressors and challenges~\cite{gable2005and,slade2010mental,waters2022positive}. Positive emotions not only improve immediate psychological states but also contribute to long-term psychological resilience and resource building, ultimately enhancing overall well-being and life satisfaction~\cite{tugade2007regulation}.

\subsubsection{Positive Computing}
In recent years, the intersection of Positive Psychology and emotion regulation has led to the emergence of various intervention strategies have been widely applied in non-clinical populations for emotion management and mental health maintenance~\cite{fredrickson2001role}. At the same time, the field of digital technology has increasingly integrated principles from positive psychology. Sander~\cite{sander2010positive} introduced the concept of ``Positive Computing'', defining it as a design approach that intentionally leverages information and communication technologies to support psychological growth, while respecting diverse understandings of well-being across individuals and communities. Building on this, the concept of positive technology was further articulated, referring specifically to the use of digital technologies to actively create and enhance positive psychological experiences for the purpose of promoting mental health~\cite{riva2012positive}. In the domain of VR, this perspective has inspired a growing body of research and practice focused on leveraging VR technologies to generate, sustain, and amplify positive emotional experiences~\cite{banos2014earth}.

\subsubsection{VR-based Emotional Regulation}
VR has demonstrated significant potential in enhancing emotion regulation~\cite{colombo2021virtual}. Existing studies have empirically validated the effectiveness of VR-based interventions in areas such as clinical psychological treatment~\cite{riva2022virtual}, emotional activation~\cite{diemer2015impact,10.1145/3629606.3629646,10494076,9792298}, immersive educational settings~\cite{9288462}, and stress management~\cite{10316434}.

Compared to traditional digital approaches to emotion regulation, VR's immersive environments help users avoid distractions caused by their physical surroundings, thereby making it easier to focus inward~\cite{10.1145/3643834.3661570}. Furthermore, VR overcomes the limitations of space and time~\cite{hamad2022virtual,singh2020significant}, and offers unique advantages in constructing personalized, immersed environments and providing innovative virtual settings for emotional expression and cognitive restructuring~\cite{zhang2024exploring}.

As VR technology gradually becomes accessible to the general public, applications such as immersive relaxation training and stress release simulations are increasingly being integrated into everyday life~\cite{10.1145/3491102.3517573}. Although digital technologies cannot completely replace professional clinical treatments, these technological innovations have opened up new pathways for implementing emotion regulation strategies~\cite{10.3389/frvir.2021.645153}. Examples include VR-based mindfulness or meditation training to reduce negative emotional states~\cite{10316433}, pain management through attentional distraction techniques within virtual environments~\cite{10765486}, and relaxation training to practice relaxation skills, thereby lowering physiological arousal~\cite{radhakrishnan2023immersive} and enabling more effective stress management~\cite{10.1145/2931002.2931017,9090656}.

\subsubsection{Relaxation in VR}
The potential of VR technology in establishing new ways of emotional expression to enhance the ability of emotional regulation, as well as in the fields of relaxation training and stress relief, has been widely verified with the popularization of VR experience systems that are becoming increasingly affordable, portable and comfortable~\cite{10.1145/3552327.3552336}. 
Most notably, modern busy lifestyles and living environments often limit opportunities for relaxation~\cite{riches2021virtual}. Unlike traditional desktop applications, which typically provide limited sensory engagement and are susceptible to distractions from the surrounding physical environment, VR technology effectively blocks external interference and overcomes temporal and spatial constraints. This allows users to become fully immersed in virtual scenes and obtain more effective relaxation experiences~\cite{10.1145/3450741.3465248,10.1145/3613905.3637139}. Additionally, personalized virtual environments have been shown to effectively relieve pain and induce relaxation~\cite{10.1145/3611659.3617214}. Empirical studies further demonstrate that VR-based relaxation consistently yields greater reductions in stress and negative affect than desktop-based interventions~\cite{liszio2018relaxing}, a benefit likely attributable to VR's heightened multisensory engagement and enhanced sense of presence~\cite{riches2021virtual}.

\subsection{Mechanisms and Applications of Affective Haptics}


The interactive actions between the body and the environment are one of the ways to express users' thoughts and emotions. Correspondingly, in the theory of embodied cognition, users' thoughts and emotions are also the result of the interaction between the body and the environment~\cite{10.1145/3613904.3642328}. The mechanism of embodied cognition determines that the brain can activate specific abstract concepts by perceiving specific movement states, provided that the connection between the movement state and the abstract concept already exists in the user's cognition~\cite{10.1145/3544549.3585805}. Touch is one of the essential channels for social interaction and emotional expression in human life. It plays an irreplaceable role in both human evolution and everyday communication. Compared to distant senses like vision and hearing, the close-range and direct nature of touch enables a deeper emotional intimacy and richer interaction depth~\cite{hertenstein2006touch}. Beyond serving as a conduit for physiological signals, touch is also considered an efficient means of emotional communication and self-regulation~\cite{hertenstein2009communication}.

In the field of human-computer interaction (HCI), researchers have increasingly explored the integration of haptic elements into interfaces to recognize, convey, or intervene in users' emotional states. Numerous studies and design practices have shown that tangible interfaces hold significant potential for enhancing emotional regulation and improving user satisfaction~\cite{10.1145/3462244.3479917,10.1145/3643834.3661608,10.1145/3064663.3064697}. On one hand, compared to conventional screen taps or button presses, tangible haptic devices, such as soft cushions, handheld stress-relief tools, or interactive plush toys, enable users to express or release emotions through diverse bodily actions like touching, hugging, squeezing, or stroking~\cite{slovak2021situ}. These devices can also deliver real-time feedback through vibrations, lighting effects, and sound, enriching the depth and authenticity of emotional communication~\cite{10.1145/3632776.3632794,7320966,yu2021vibreathe}. On the other hand, by embedding sensors or actuators into physical devices, systems can collect real-time data on users' tactile interactions and map these discrete inputs to corresponding emotional indicators, offering technical support for emotion-aware feedback and intervention~\cite{10.1145/3643834.3661608,zhou2017textile}.

During haptic interaction, people often use gestures such as stroking, hugging, patting, gripping, or squeezing to convey or regulate emotions, expressing care, relieving stress, or venting anxiety~\cite{mcdaniel2019therapeutic,ullan2014effect,10812211}. These behaviors are typically spontaneous and emotionally meaningful~\cite{10.1145/2399016.2399134,hunter1997ethical}. Among these, squeezable interaction is particularly well-suited to emotional expression and regulation~\cite{10.1145/3623509.3635256,10.1145/3462244.3479917,10.1145/3706598.3713483}, making it a valuable modality in affective interaction design and the specific form of squeeze interaction adopted in this study.

First, squeezing is one of natural interaction that people often instinctively apply pressure with their hands to release emotional tension when experiencing stress, anxiety, or mood fluctuations, an action characterized by universality and spontaneity~\cite{10.1145/2399016.2399134}, which shows it is not merely a mechanical act, but a deeply embodied practice that links bodily movement with emotional and cognitive processing~\cite{wilson2002six,barsalou2008grounded}. Moreover, the squeezable design provides users with a sense of control~\cite{10.1145/3010915.3010930}. In a VR environment, manipulating a tangible interface that offers physical feedback and reproduces character actions can enhance the sense of presence and strengthen the connection between the user and the virtual character (sense of embodiment).~\cite{10322238}. Second, squeeze gestures are easily quantifiable in real time via pressure sensors, allowing the system to accurately capture key parameters such as intensity, frequency, and duration of the applied force, thus enabling objective analysis and identification of users' emotional states and trends~\cite{10.1145/2501988.2502033}.

\section{System Design}
To achieve immersion, interactivity, and effectiveness in the VR Calm Plus experience, we developed a comprehensive multimodal design framework that integrates hardware and software systems, interaction mechanisms, and content scripts. Centered around user perception, this design systematically maps squeeze-based input to visual and auditory feedback. It also defines the usage patterns, data logic, interaction content, and technical implementation pathways. In this section, we provide a detailed explanation of the design.

\subsection{Overview of VR Calm Plus}
The design of VR Calm Plus draws inspiration from the Mechanics-Dynamics-Aesthetics (MDA) framework~\cite{hunicke2004mda}, a widely adopted approach for conceptualizing user experience in interactive systems and digital games. The MDA framework articulates user experience as the interplay among three hierarchical components: mechanics (the system's rules and logic), dynamics (the run-time behaviors that emerge from user-system interaction), and aesthetics (the resulting emotional responses and experiential qualities).  Building upon this foundation, we applied the MDA framework to the design of a VR-based emotion regulation system, systematically integrating system functionalities, interaction modes, and target emotional outcomes.

(1) At the mechanics level, we define concrete rules and mappings between haptic input (e.g., squeezing the plush device) and real-time virtual feedback (such as visual, auditory, and ambient environmental changes). These rules are designed to translate users' physical actions into meaningful changes in the virtual environment, providing immediate, perceivable feedback that supports emotion regulation through bodily engagement and multisensory cues. In our study, we implement two conditions: (a) \textbf{Squeeze Interaction} condition, in which participants hold the plush device and their squeezing input actively controls the VR mappings; and (b) \textbf{Audio-Visual} condition, in which participants do not hold the plush and therefore provide no physical input to the system. This design isolates the effect of active squeeze interaction engagement from the absence of physical interaction.

(2) At the dynamics level, the focus is on the temporal and adaptive aspects of interaction, capturing how users' actions and emotional states evolve during the session. The system monitors and responds to users' ongoing squeezing inputs, modulating feedback intensity and scenario progression accordingly. This dynamic loop aims to foster a sense of agency and emotional co-regulation, enabling users to transition through various emotional phases, such as adaptation, active guidance, and eventual relaxation.

(3) At the aesthetics level, the design emphasizes the creation of a rich, immersive, and emotionally resonant atmosphere. This is achieved through carefully orchestrated scene transitions, gradual visual fades, integrated music cues, and verbal prompts, all intended to evoke feelings of safety, calmness, and emotional release. 

The resulting design incorporates three interactive scenarios based on targeted experiential and affective goals, with specific design considerations outlined below. These three scenarios represent a progressive journey through the stages of ``Adaptation and Shaping'', ``Interaction and Guidance'', and ``Meditation and Relaxation''. Transitions between these three stages rely on scene switching, gradual fades, integrated music cues, and verbal prompts, ensuring users remain immersed throughout the experience. As the narrative tempo (time frames) progresses, users pass through different emotional phases: from adaptation to active control, and finally to emotional release.

\begin{figure*}[htbp]
    \centering
    \begin{minipage}[t]{0.42\textwidth}
        \centering
        \begin{subfigure}[t]{\linewidth}
            \centering
            \includegraphics[width=\linewidth]{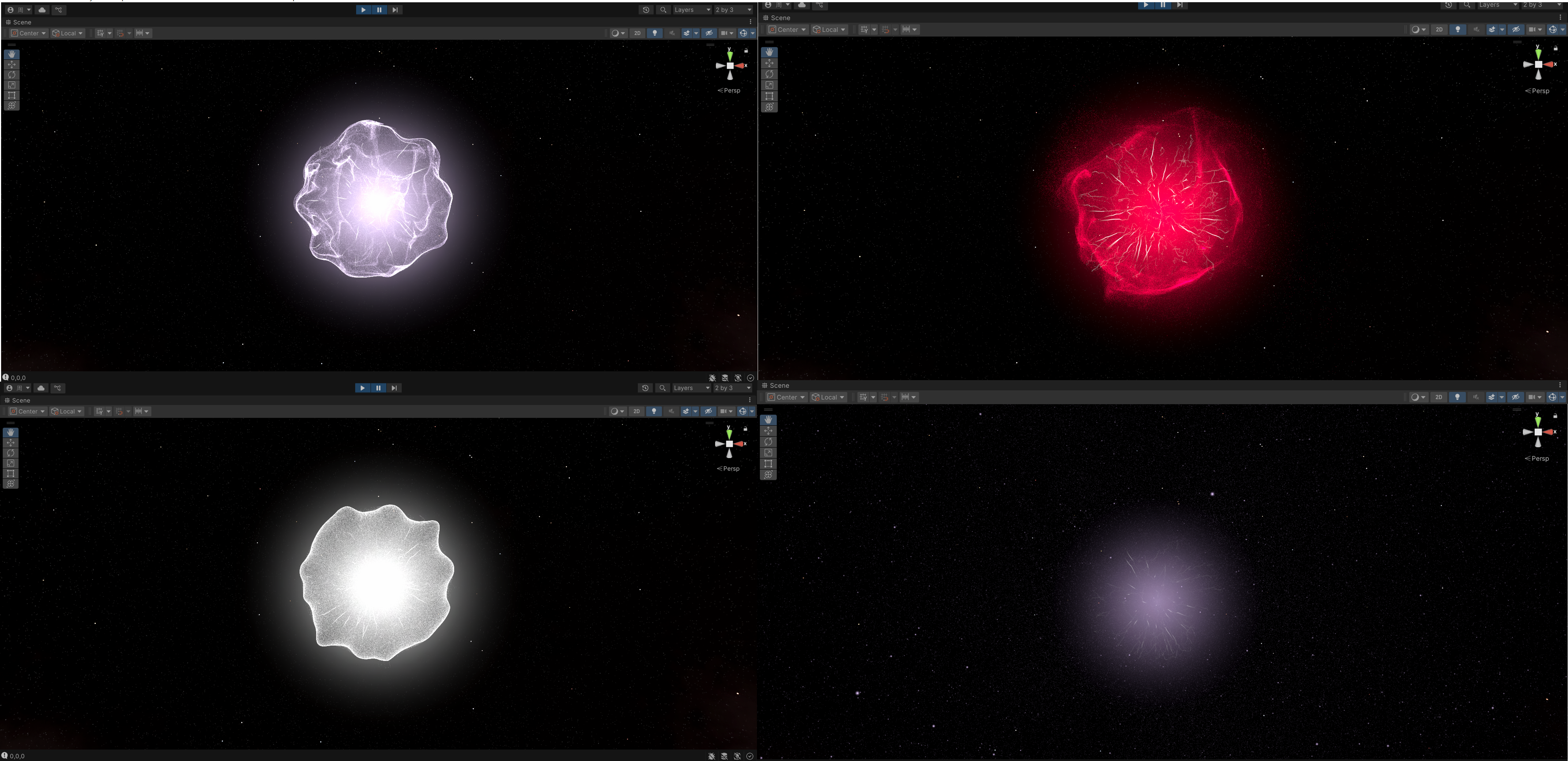}
            \caption{``Adaptation and Shaping'' Stage}
            \label{fig:s1}
        \end{subfigure}
        \vspace{1em}
        \begin{subfigure}[t]{\linewidth}
            \centering
            \includegraphics[width=\linewidth]{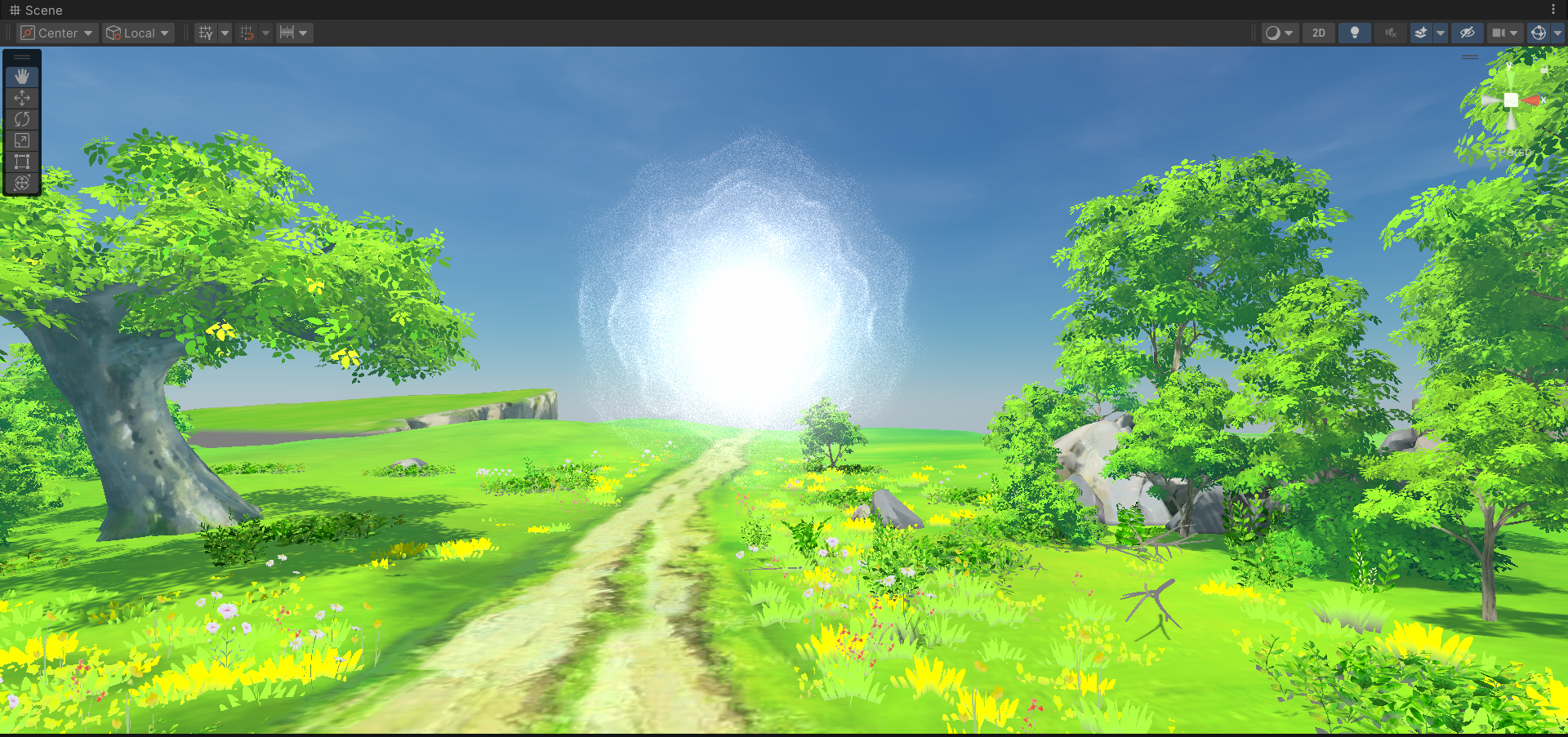}
            \caption{``Interaction and Guidance'' Stage}
            \label{fig:s2}
        \end{subfigure}
    \end{minipage}
    \hspace{1em}
    \begin{minipage}[c]{0.5\textwidth} 
        \centering
        \begin{subfigure}[t]{\linewidth}
            \centering
            \includegraphics[width=\linewidth]{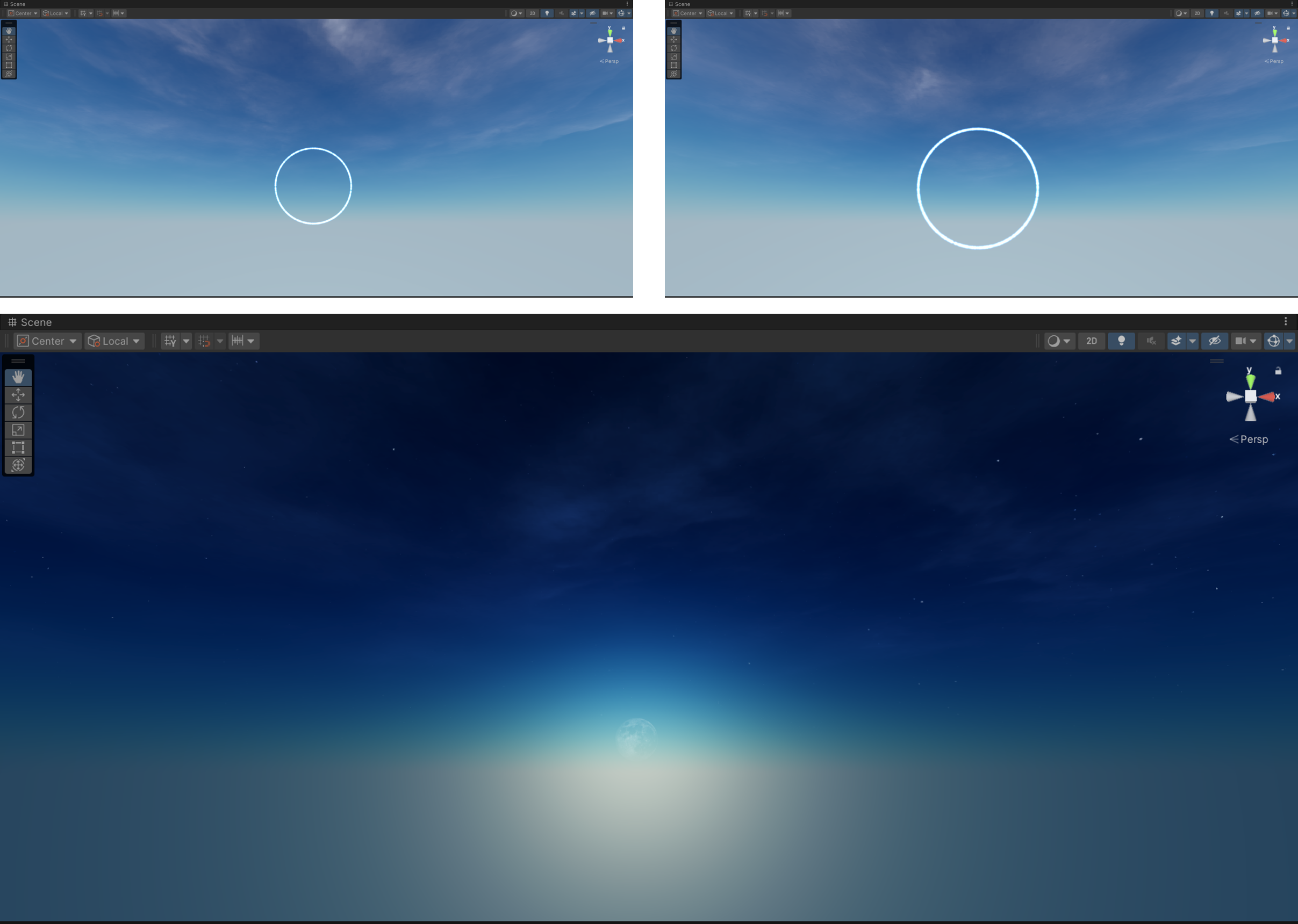}
            \caption{``Meditation and Relaxation'' Stage}
            \label{fig:s3}
        \end{subfigure}
    \end{minipage}
    \vspace{-0.2cm}
    \Description{VR Calm Plus consists of three stages. In the Adaptation and Shaping stage, participants are guided to calm down and become familiar with the interaction methods and particle effects. In the Interaction and Guidance stage, they release their emotions and engage in a large number of interactions. In the Meditation and Relaxation stage, they gradually gathered their emotions until they calmed down.}
    \caption{Overview of the three stages in VR Calm Plus: 
    (a) Adaptation and Shaping, showing vibrant particle-based orbs against a star-filled background; 
    (b) Interaction and Guidance, presenting an expansive green field and a glowing white orb that guides attention through foreground, mid-ground, and background; 
    (c) Meditation and Relaxation, depicting a dynamic white circle and a transition from clear sky to starry horizon that evokes calm and reflection. 
    Together, these stages illustrate the progressive shift from energetic visualizations toward tranquil, immersive atmospheres.}
    \label{fig:vr_calm_stages}
\end{figure*}
\subsubsection{1st Scenario - Adaptation and Shaping - Stage 1}
The first scenario, referred to as the Adaptation and Shaping stage, aims to help users gradually acclimate to the immersive VR environment and become familiar with the interaction methods. The emotional regulation goal is to give users a sense of agency over their emotional state. Users interact with a tangible interface by applying varying levels of pressure, primarily leveraging the squeeze-based haptic modality. This pressure influences the speed of particle motion, particle color transitions, and the tempo of background music, thereby fostering user awareness and control over active emotional regulation.

To establish a robust sense of agency, we prioritized predictability and temporal contiguity in the interaction design~\cite{haggard2017sense}. Although a single input controls multiple feedback modalities (particle speed, color, and musical tempo), this one-to-many mapping was deliberately designed to be intuitive and instantaneous. Specifically, we implemented a direct linear mapping strategy (raw pressure $0$--$80,000$) without artificial smoothing. As the user applies pressure, the particle motion accelerates, and the color transitions smoothly from cool white to saturated purple and finally to warm red, accompanied by a linearly modulated musical pitch ($0.9\times$--$1.1\times$). This immediate, predictable responsiveness ensures that the virtual environment reacts as a direct extension of the user's body, minimizing the cognitive load of operation while maximizing the feeling of control~\cite{limerick2014experience}. Aesthetically, a starry sky and dynamic particle spheres serve as core elements to create a vast, ethereal, and immersive atmosphere (see Figure~\ref{fig:s1}).

\vspace{-0.2cm}
\subsubsection{2nd Scenario - Interaction and Guidance - Stage 2}
In the second scenario, the Interaction and Guidance stage, VR Calm Plus focuses on further elevating users' emotional engagement and actively guiding them toward a deeper experience. The emotional regulation goal is to fully capture users' sensory attention, increasing emotional arousal and concentration. By transitioning users' emotional state from a low-arousal zone to a more enjoyable, high-arousal zone, this stage aims to foster greater interactive enthusiasm and exploratory interest.

Technically, this stage introduces real-time audio spectrum analysis to drive the rhythmic fluctuations of the particle system. Unlike the first scenario where user input modulated the music, here the music acts as a fixed auditory guide. We selected a track with a steady tempo of 108 BPM and a 4/4 time signature to establish a clear rhythmic foundation. Users are encouraged to synchronize their squeezing actions with the strong beats (the 1st and 3rd beats of each measure). To prioritize relaxation over precision, we implemented a tolerant target window of $\pm 300$ ms around the beat. Successful synchronization triggers a distinct feedback signal: the particles pulse outward and brighten, providing immediate positive reinforcement. In this mode, the user's squeeze intensity is mapped solely to the particle color gradient (White--Purple--Red), while the music tempo remains constant. As the pressure level shifts from weak to strong, the particle color gradually moves from cooler tones to warmer tones (see Figure~\ref{fig:s2}). This close interplay of auditory, visual, and squeeze-based haptic modalities helps users quickly enter a positive, engaged state.



\subsubsection{3rd Scenario - Meditation and Relaxation - Stage 3}
In the third scenario, the Meditation and Relaxation stage, the goal is to facilitate deeper relaxation and self-awareness, guiding users back to a low-arousal, pleasant, and calm emotional state. This completes the emotional regulation loop within the overall experience. Visual cues, such as a tranquil waterscape merging with the sky, take the lead in this scenario, supplemented by interactions (low-frequency squeeze feedback) and auditory elements (meditative music). Accompanied by guided breathing exercises, presented through a dynamically contracting circle that prompts users to squeeze and breathe at a steady pace, the focus is on breathing awareness and bodily perception, enabling further emotional regulation. Unlike adaptive biofeedback, this guide follows a fixed exogenous pacing protocol (6 breaths/minute: 4s inhalation, 6s exhalation) to establish a stable rhythm known to enhance vagal tone~\cite{lehrer2014heart}. (see in Figure~\ref{fig:s3}.)

\vspace{-0.3cm}

\subsection{Squeeze-based Haptic Device Integration}
Squeeze interaction is one of the core components of VR Calm Plus. To provide a natural and intuitive method for users to engage with the virtual world, we integrated a physical squeeze-based haptic interface within the VR environment. Specifically, we employed two flexible force-sensitive resistor (FSR) sensors, placed on the front and back surfaces of a plush toy, ensuring that any change in grip pressure can be effectively detected regardless of how the user holds the toy. To enhance the stability and accuracy of data collection, we used two independent linear voltage conversion modules (see Figure 4). Each FSR sensor is connected to a separate module input; the modules then deliver a stable 0–5V analog voltage output, which is transmitted to the Arduino Mega's analog inputs A0 and A1. Both modules are powered by the board's 5V and GND pins to ensure reliable signals. The Arduino board communicates with the computer via USB serial connection, sending processed data to Unity in real time.

\subsubsection{Plush Toy Embedding Structure Design}
To balance user comfort and effective interaction, we selected a spherical plush toy as the physical interface. A small zipper was sewn onto the back of the toy, allowing us to embed the two FSR sensors on its front and back interior surfaces. Cotton stuffing was added to protect the sensors and enhance haptic comfort. The sensor wires run from inside the toy through the zipper opening and connect to the linear voltage conversion modules and the development board. Finally, the zipper is closed to keep the toy's exterior neat and minimize any extraneous wiring that could interfere with user interaction.

\subsubsection{Interaction Mapping and Feedback Mechanisms}
To achieve real-time linkage between user squeezing actions and visual or auditory feedback, we establish a stable communication channel between hardware and software, alongside responsive visualization and control logic. Our VR Calm Plus system embeds two FSR sensors in a spherical plush toy to detect changes in grip intensity, transmitting the resulting data to Unity through an Arduino. This pipeline ensures immediate, synchronized feedback via direct linear mapping: raw pressure input ($0$-$80,000$) is instantaneously interpolated to drive particle color gradients (White-Purple-Red) and modulate background music pitch ($0.9\times$-$1.1\times$). To prioritize responsiveness and the user's sense of agency, no artificial software smoothing was applied, resulting in minimal system latency determined primarily by the hardware serial communication rate. Additional technical details regarding hardware-software data communication, data visualization and real-time feedback, and the control of scene progression and interactions can be found in the Supplemental Material.

\vspace{-0.7em}

\section{Methods and Experimental Setup}

\subsection{Participants Recruitment}

We recruited participants through social media flyers and snowball sampling, targeting individuals over 18 years old. All potential participants voluntarily completed an application questionnaire that included demographic information, a psychological assessment scale (State Anxiety Inventory~\cite{spielberger1983state}), and contact details. We screened these applications and invited respondents to participate in our study on a first-come, first-served basis. During screening, we excluded applicants who did not meet our requirements (minors, people with conditions unsuitable for VR experiences such as heart disease, hypertension, and motion sickness, as well as duplicate submissions, incomplete questionnaires, and forms containing invalid information). We also balanced gender distribution among participants. Our final sample consisted of 40 participants (22 females, 18 males, average age $=$ 28 years old, SD $=$ 12.03) who completed the field experiment (see in Table~\ref{table:demograph_summary} and \ref{table:demograph_participants} for full demographic information). The headset (Meta Quest 3) used in this study is designed to accommodate users wearing glasses. Consequently, while we excluded individuals with specific eye diseases, we did not restrict participants who wear glasses. Notably, we received no reports (explicit or implicit) of discomfort related to wearing glasses during the study. The study was approved by the university's Institutional Review Board (IRB) (ID: anonymous). To compensate participants for their time (approximately 45–50 minutes), each participant received a cash reward of 50 Chinese Yuan upon completion of the study.

\begin{table}[htbp]
\centering
\caption{Demographic profile of study participants ($N=40$).}
\vspace{-0.7em}
\label{table:demograph_summary}
\small
\renewcommand{\arraystretch}{1.2}
\begin{tabular}{llcc}
\toprule
\textbf{Category} & \textbf{Sub-category} & \textbf{Count} & \textbf{Percentage (\%)} \\
\midrule
\multirow{2}{*}{\textbf{Gender}} 
 & Female & 22 & 55 \\
 & Male & 18 & 45 \\
\midrule
\multirow{2}{*}{\textbf{Age}} 
 & Mean (SD) & \multicolumn{2}{c}{28 (12.03)} \\
 & Range & \multicolumn{2}{c}{19 -- 66} \\
\midrule
\multirow{5}{*}{\shortstack[l]{\textbf{Meditation}\\\textbf{Experience}}}
 & Never & 14 & 35.0 \\
 & Initial Exposure & 11 & 27.5 \\
 & Basic Understanding & 8 & 20.0 \\
 & Experienced & 3 & 7.5 \\
 & Proficient & 4 & 10.0 \\
\midrule
\multirow{4}{*}{\textbf{VR Experience}} 
 & Never & 9 & 22.5 \\
 & Initial Exposure & 14 & 35.0 \\
 & Basic Understanding & 13 & 32.5 \\
 & Proficient & 4 & 10.0 \\
\bottomrule
\end{tabular}
\end{table}

\vspace{-0.7em}
\subsection{Experimental Environment Setup}

The experiment was conducted in a dedicated room, with only one participant and two researchers present during each session. The participant remained seated throughout the experiment. Three computers were used in the study. One computer was connected to the VR device (Quest 3) and the device of squeeze interaction used by the participant, handling the synchronization, playback, and recording of the content. The other two computers were used by the researchers to monitor the entire experimental process, record the study procedure. Additionally, a camera mounted on a tripod was set up in the room to record the entire process of each experiment. To improve the efficiency and convenience of completing questionnaires, the researchers generated a unique QR code for each questionnaire and placed it next to the participant's seat, enabling the participant to complete them using either their own device or the researchers' devices.
\vspace{-0.7em}
\subsection{Data Collection, Measures and Analysis Methods}
To rigorously evaluate user experiences under different experimental conditions (Squeeze Interaction vs. Audio-Visual), we adopted quantitative methods. This subsection details our procedures for data collection, utilized measures, and analytical techniques.
\vspace{-0.7em}
\subsubsection{Data Collection}

We collected multimodal data from 40 participants across all VR sessions. Participants' affective states were assessed using standardized questionnaires at three distinct points: pre-test, mid-test, and post-test. Additionally, subjective evaluations of VR experiences were gathered through custom-designed questionnaires administered immediately after each VR session. 
Simultaneously, objective physiological data were continuously recorded throughout the experiments using a wearable biosensing wristband (Psychorus, Huixin, China). This multimodal approach allowed us to corroborate subjective reports with real-time physiological evidence.

\subsubsection{Measures}

\paragraph{PANAS-X scales:} The Positive and Negative Affect Schedule–Expanded (PANAS-X) scales~\cite{watson1994panas} were utilized as standardized measures to systematically assess participants' emotional states before, during, and after the VR sessions. These scales provided a reliable means of detecting affective changes resulting from each experimental condition~\cite{10.1145/3629606.3629646}.

\paragraph{Custom Subjective Experience Evaluation Questionnaire:} 
We developed a custom subjective evaluation questionnaire comprising 16 universal questions, designed to assess participants' experiences across both Squeeze Interaction and Audio-Visual conditions. All items were rated on a 5-point Likert scale (1 = ``strongly disagree / very poor'' to 5 = ``strongly agree / very good''). The questionnaire focused on five core dimensions: immersion, emotional regulation, scene design, interaction clarity, and overall user satisfaction. Identical questions were used across both conditions to ensure consistency and comparability.

\paragraph{Physiological Apparatus:}
To capture real-time physiological responses without hindering hand movements, we utilized a wireless biosensing wristband (Psychorus, Huixin, China). This device integrates a multi-modal sensor array, primarily consisting of an optical Photoplethysmography (PPG) sensor and dry electrodes for Electrodermal Activity (EDA) sensing. The PPG sensor emits green light to detect blood volume changes in the microvascular bed, from which we derived Heart Rate (HR) and Pulse Rate Variability (PRV) to index physiological arousal and autonomic regulation. Simultaneously, the EDA sensors measured skin conductance (galvanic skin response, GSR) via two electrodes in contact with the wrist, serving as an indicator of emotional intensity. An internal Inertial Measurement Unit (IMU) monitored wrist motion to facilitate artifact removal.

\subsubsection{Questionnaires' Analysis Methods}

We conducted statistical analyses on the data from the PANAS-X scales and the subjective experience questionnaires to identify differences in emotional responses and user ratings between the conditions. Given the cross-over experimental design, standard statistical tests, including paired-samples t-tests, were employed to assess within-subject differences. For the PANAS-X, in addition to conducting a statistical analysis on each individual emotion score, we also merged the more fine-grained emotion labels into broader categories according to Watson and Clark's classification approach~\cite{watson1994panas}.

\subsubsection{Physiological Data Analysis}
Physiological data processing followed a rigorous multi-stage pipeline implemented in Python. First, the raw time-series data from the wristband were synchronized with the VR event logs to precisely segment the data into three phases: Baseline, Audio-Visual, and Squeeze Interaction. To ensure data integrity, we applied a two-step noise reduction strategy. We first employed a \textit{physiological range filter} to eliminate hardware artifacts, discarding biologically implausible values (e.g., Heart Rate $<$ 40 or $>$ 180 bpm; GSR $\le$ 0.01 $\mu$S). Subsequently, a statistical filter based on the Interquartile Range (IQR) was applied. Data points falling outside the range $[Q1 - 1.5 \times IQR, Q3 + 1.5 \times IQR]$ within each condition were identified as outliers and removed. This step was crucial for preserving genuine high-arousal responses while filtering out transient noise caused by body movements. Given the inherent inter-subject variability in physiological baselines, analyzing absolute values could yield misleading results. Therefore, we normalized the data by calculating the ``Change from Baseline'' ($\Delta$) for each metric. For a given metric $M$ in an experimental condition, the delta value was calculated as $\Delta M = M_{raw} - \mu_{baseline}$, where $\mu_{baseline}$ is the participant's mean value during the resting baseline phase of that specific session. Finally, we aggregated these $\Delta$ values and conducted independent samples t-tests to compare the physiological effects between the Squeeze Interaction and Audio-Visual conditions.

\subsection{Procedure}

\begin{figure}[htbp] 
    \centering
    \includegraphics[width=1\columnwidth]{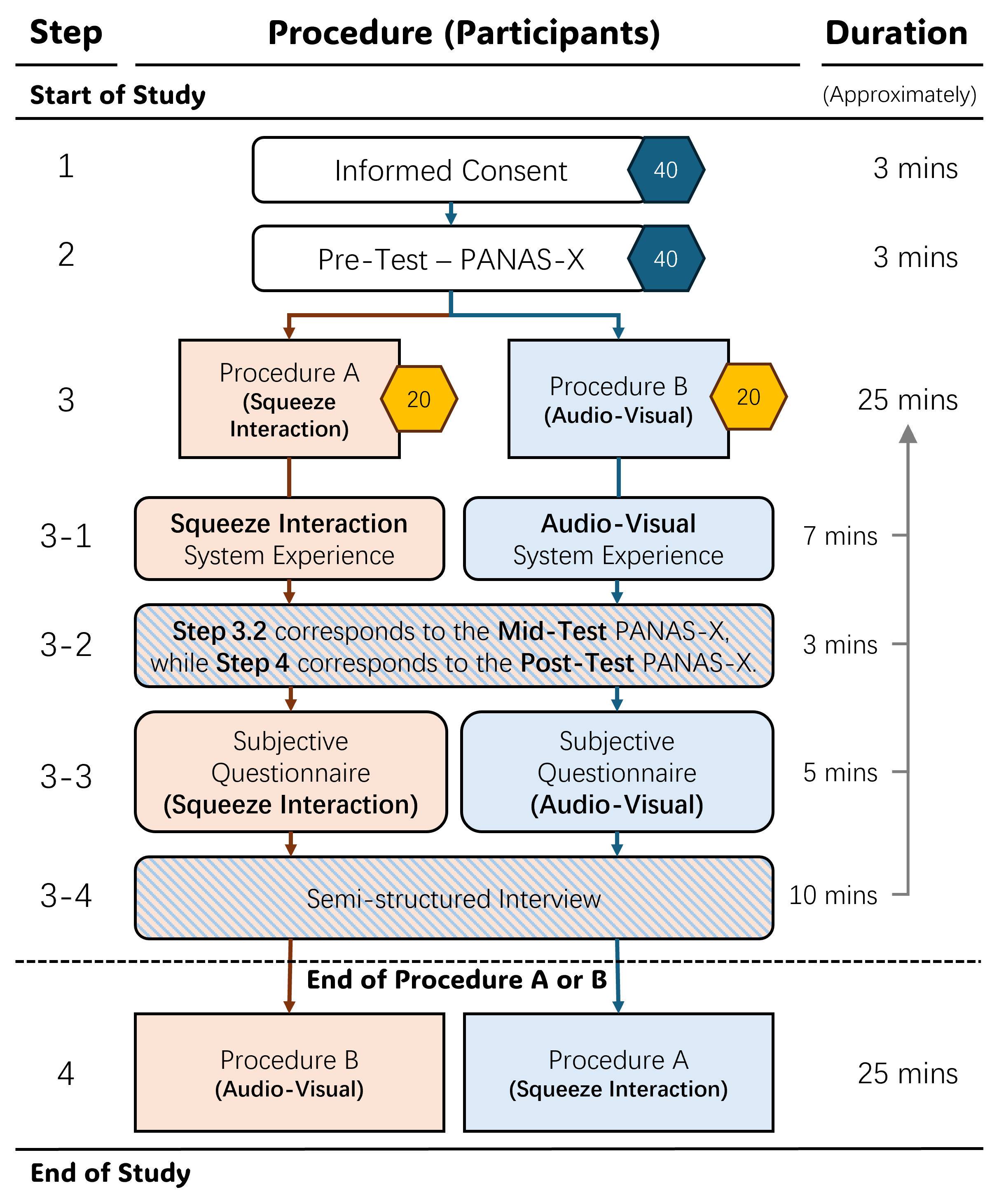}
    \Description{Forty eligible participants were spontaneously divided into groups A and B according to the available time periods, with 20 people in each group. Group A, numbered 1 to 20, experience the Squeeze Interaction‐enabled VR system first. Group B, numbered 21 to 40, experience the Audio-Visual vision first.}
    \caption{Overview of the Experimental Process and Research Design. The process flows from top to bottom, and the content in Step 4 is identical to that in Step 3, with the only distinction being between Squeeze Interaction and Audio-Visual experiences, i.e., Procedure A (Squeeze Interaction→Audio-Visual): participants experienced the Squeeze Interaction‐enabled VR system first, then the Audio-Visual version, and Procedure B (Audio-Visual→Squeeze Interaction): participants experienced the Audio-Visual version first, then the Squeeze Interaction version.}
    \label{fig:procedure}
\end{figure}

We employed a within‐subjects cross‐over design~\cite{privitera2024research} in which each participant completed two sessions. Figure~\ref{fig:procedure} illustrates the overall flow of our within‐subjects, counterbalanced cross‐over study with 40 participants.  We divided participants into two equal groups. One with squeeze feedback and one without. To counterbalance order effects, half of our participants (N = 20) experienced the Squeeze Interaction condition first (Procedure A), and the other half (N = 20) began with the Audio-Visual condition (Procedure B). All participants completed a baseline PANAS-X questionnaire (Pre-Test) prior to any interaction. Following each condition, participants completed a post- PANAS-X questionnaire and a custom subjective experience questionnaire. This counterbalanced design allowed us to evaluate both main effects of condition and interaction effects between condition and order.

Each participant completed two sessions (Squeeze Interaction and Audio-Visual), comprising ten steps in total. The detailed descriptions and procedures for each step are as follows:
\begin{itemize}
    \item \textbf{Informed Consent ($\sim$3 mins).}  
    We provided a printed consent form, explained the study purpose and procedures, and collected signed consent from all participants.

  \item \textbf{Pre‐Test – PANAS‐X ($\sim$3 mins).}  
    Participants scanned a QR code and completed the Positive and Negative Affect Schedule (PANAS‐X) on their mobile device to record baseline affect.

  \item \textbf{VR System Experience ($\sim$7 mins).}  
    For the Squeeze Interactions' session, we fitted the VR headset, physiological sensors, and squeezable device, then gave a brief tutorial before participants explored three sequential VR scenes using the VR Calm Plus interface.  
    For the Audio-Visual session, the same procedure was followed without activating the squeezable device.

  \item \textbf{Mid/Post‐Test – PANAS‐X ($\sim$3 mins).}  
    After each VR experience, participants immediately scanned a QR code to complete the PANAS‑X again, capturing their most recent affective changes.

  \item \textbf{Subjective Questionnaire ($\sim$5 mins).}  
    Participants rated seven dimensions of overall system experience (immersion, enjoyment, control, usability, emotional resonance, relaxation, emotion regulation) and three scene‐specific items using a 5‐point Likert scale.  
    In the Squeeze Interactions' session, they also completed a five‐item questionnaire (naturalness, coordination, immersion enhancement, emotional expressivity, comfort).

\item \textbf{Interview ($\sim$10 mins).}  
    We conducted a semi‐structured interview covering overall impressions, emotional changes, interaction feedback, scene‐by‐scene reflections, and any additional comments.

  \item \textbf{Cross‐Over.}  
    After participating in the first session of the study, participants switched to the other condition (Squeeze Interaction or Audio-Visual) and repeated steps from 3-1 to 3.3 for the second session (step 4).
\end{itemize}

\section{Results}
\subsection{Quantitative Findings}

\subsubsection{\textbf{PANAS-X Scales}}
\begin{figure*}[htbp] 
    \centering
    \includegraphics[width=1\linewidth]{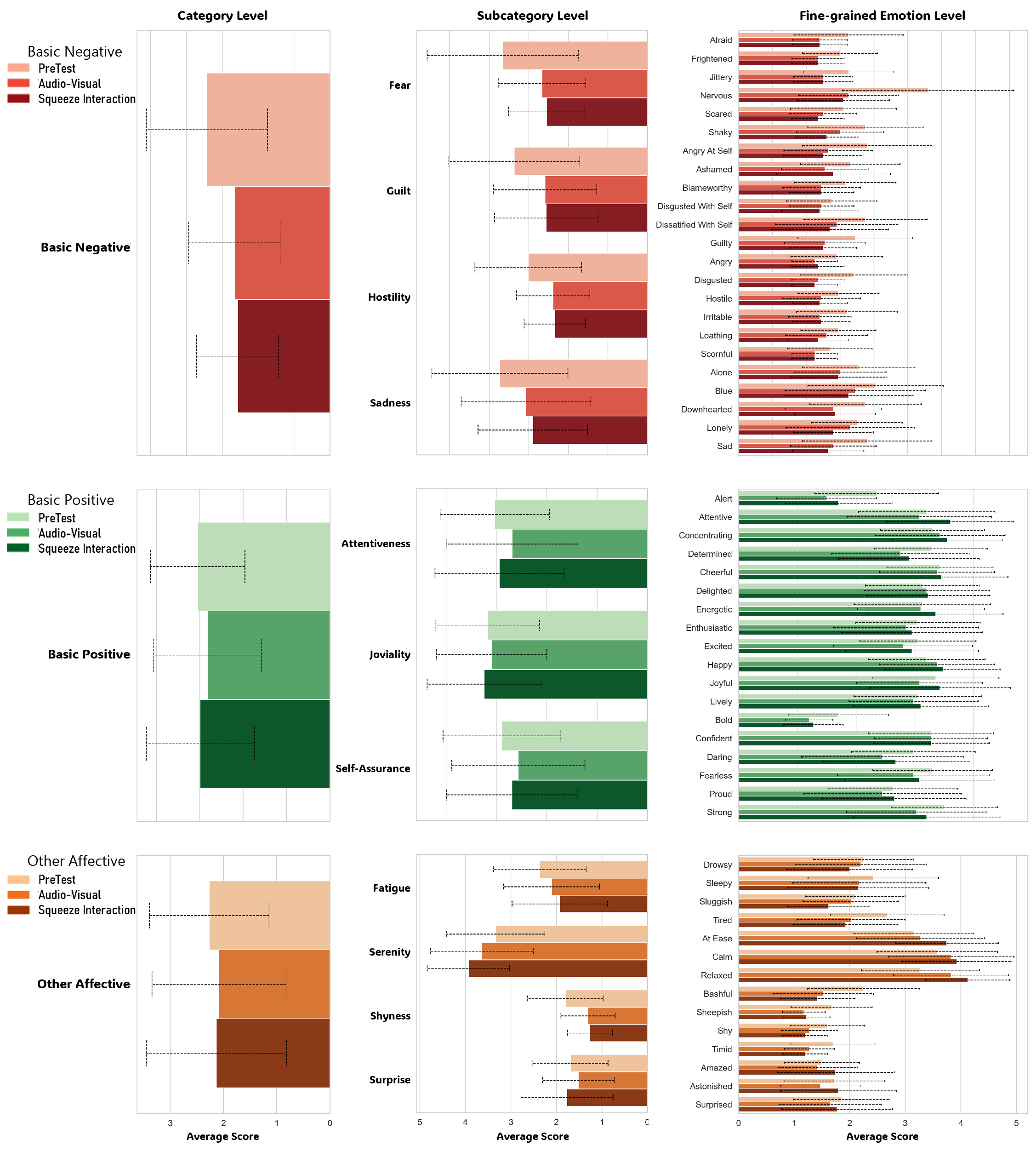}
    \Description{We collected PANAS-X data of 40 participants in three types: ``before experience'', ``after Squeeze Interaction version'', and ``after Audio-Visual version'' for each person, and calculated the average value of each item for 40 people under each of the three type.}
    \caption{Hierarchical overview of emotional responses across experimental conditions (PANAS-X). This visualization displays the mean emotional scores, stratified by three levels of granularity: Category (left, e.g., Basic Positive), Subcategory (middle, e.g., Serenity), and Fine-grained Emotion (right, e.g., Relaxed). The data is compared across three conditions: PreTest (lightest shade), Audio-Visual (medium shade), and Squeeze Interaction (darkest shade). The conditions are distinguished by varying contrast within the same hue family for each category (Middle/Green for Positive, Upper part/Red for Negative, Bottom part/Orange for Other Affective), providing high contrast for accessibility. The Squeeze Interaction's condition generally elicited higher positive emotional responses, particularly in dimensions of Serenity and Attentiveness, while maintaining low negative affect comparable to baseline. Error bars represent standard deviations.}
    \label{fig:barall}
\end{figure*}

The Figure~\ref{fig:barall} shows the average scores for each of the 55 fine-grained emotional states (grouped by subcategories and higher-level categories) across the three experimental conditions: Pre-Test, Squeeze Interaction and Audio-Visual. Overall, after experiencing VR Calm Plus, most emotions across different conditions exhibited similar trends: negative emotions (e.g., ``nervous,'' ``sad,'' ``disgusted,'' ``angry at self'') were notably reduced, while positive emotions (e.g., ``happy,'' ``relaxed,'' ``calm'') showed a clear increase. When comparing the Squeeze Interaction and Audio-Visual conditions, the Squeeze Interaction experience generally produced more positive effects and led to a stronger decrease in negative emotions. Some emotions showed especially large shifts in their scores, including positive states such as ``relaxed,'' ``concentrating,'' ``at ease,'' and ``calm,'' along with negative states like ``blue,'' ``dissatisfied with self,'' ``drowsy,'' ``lonely,'' and ``shaky.'' These findings suggest that the Squeeze Interaction experience further promotes relaxation, focus, reassurance, and calmness, and helps reduce users' negative affect. We also observed that for certain emotional measures, the Squeeze Interaction condition showed distinct advantages. For instance, ``joyful'' decreased under the Audio-Visual condition compared to the pre-test but increased under the Squeeze Interaction condition. Similar differences appeared in ``lively,'' ``amazed,'' ``attentive,'' ``cheerful,'' and ``energetic,'' indicating the unique benefits of incorporating squeeze interaction feedback.

To evaluate the emotional impact of squeeze interaction feedback, we conducted paired-sample t-tests comparing participants' self-reported emotion scores between the Squeeze Interaction and Audio-Visual conditions across eleven summarized PANAS-X emotion dimensions.

\begin{table}[htbp]
\centering
\small
\Description{Statistical comparison of emotional states between Squeeze Interaction and Audio-Visual conditions. Attentiveness, Serenity, and Surprise show significant differences with medium effect sizes.}
\caption{Through paired t-tests, the significant impact of Squeeze Interaction and Audio-Visual on the emotional states of the following 11 dimensions was determined. Effect sizes (Cohen's \textit{d}) and 95\% Confidence Intervals (CI) are included.}
\label{tab:ttest_results}
\scalebox{0.85}{
\begin{tabular}{lrrccr}
\toprule
\textbf{Emotion} & \textbf{Mean Diff} & \textbf{\textit{t}} & \textbf{Cohen's \textit{d}} & \textbf{95\% CI} & \textbf{\textit{p}-value} \\
\midrule
\textbf{Attentiveness (4)} &  0.25 &  2.37 & 0.38 & $[~0.05, 0.70~]$ & .0229$^{*}$ \\
Fatigue (4)       & -0.18 & -1.83 & -0.29 & $[-0.61, 0.03]$ & .0744\phantom{$^{*}$} \\
Fear (6)          & -0.06 & -1.62 & -0.26 & $[-0.57, 0.06]$ & .1140\phantom{$^{*}$} \\
Guilt (6)         & -0.02 & -0.61 & -0.10 & $[-0.41, 0.21]$ & .5438\phantom{$^{*}$} \\
Hostility (6)     & -0.02 & -0.48 & -0.08 & $[-0.39, 0.23]$ & .6316\phantom{$^{*}$} \\
Joviality (8)     &  0.15 &  1.47 & 0.23  & $[-0.08, 0.55]$ & .1502\phantom{$^{*}$} \\
Sadness (5)       & -0.08 & -1.45 & -0.23 & $[-0.54, 0.08]$ & .1547\phantom{$^{*}$} \\
Self-Assurance (6)&  0.13 &  1.72 & 0.27  & $[-0.04, 0.59]$ & .0925\phantom{$^{*}$} \\
\textbf{Serenity (3)}      &  0.29 &  2.25 & 0.36  & $[~0.04, 0.68~]$ & .0303$^{*}$ \\
Shyness (4)       & -0.05 & -1.19 & -0.19 & $[-0.50, 0.13]$ & .2430\phantom{$^{*}$} \\
\textbf{Surprise (3)}      &  0.26 &  2.46 & 0.39  & $[~0.07, 0.71~]$ & .0185$^{*}$ \\
\bottomrule
\multicolumn{6}{l}{\footnotesize{$^{*}$Significant at $p < .05$. CI = Confidence Interval.}}
\end{tabular}
}
\vspace{-1.0em}
\end{table}
\begin{figure*}[htbp] 
    \centering
    \includegraphics[width=0.8\linewidth]{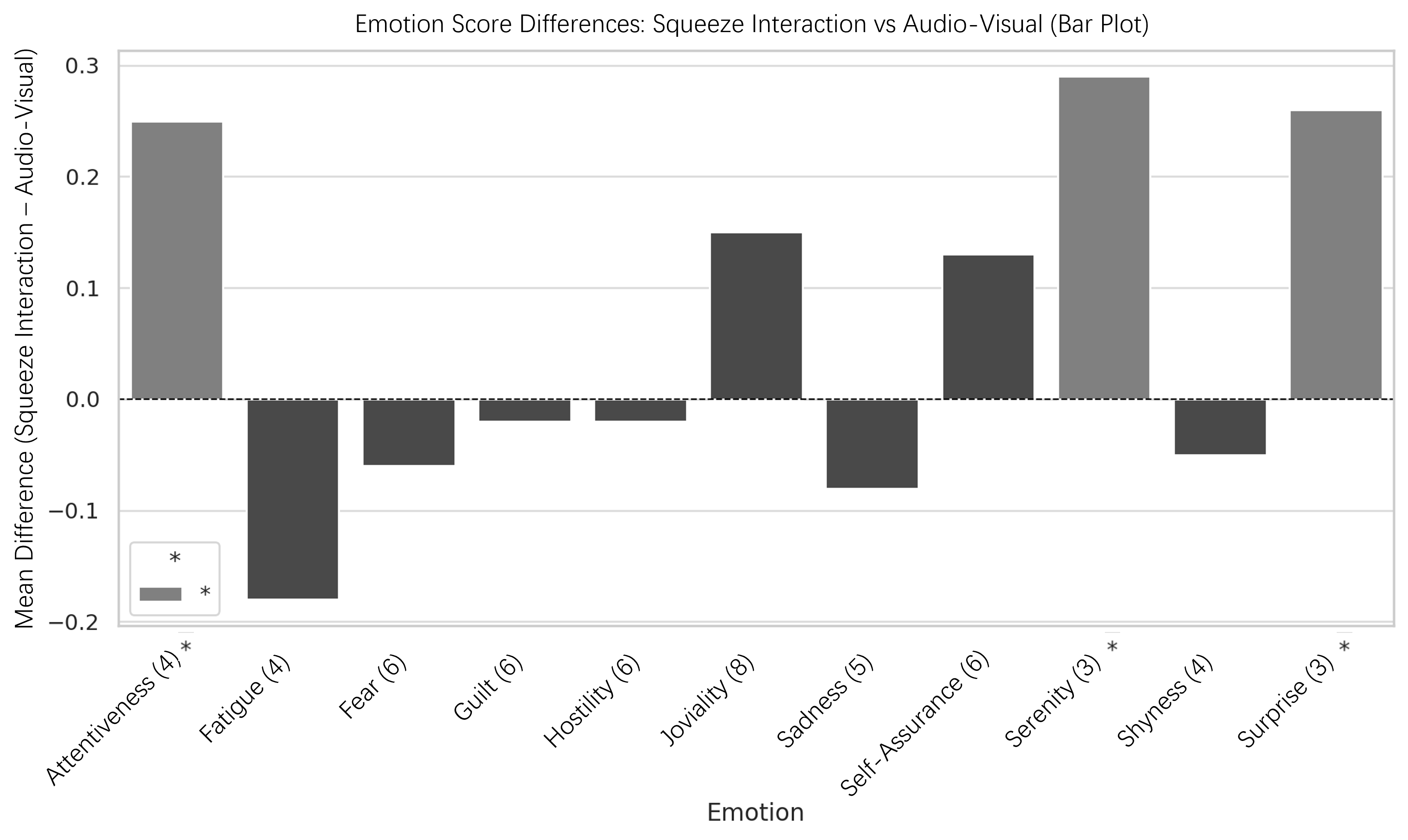}
    \vspace{-1.0em}
    \Description{Emotional scores are significantly higher under the squeeze condition for Attentiveness, Serenity and Surprise. In contrast, the score for Fatigue is relatively higher under the Audio-Visual condition.}
    \caption{Emotion Score Differences between Squeeze Interaction and Audio-Visual. The bar plot illustrates the average difference in emotional scores between the Squeeze Interaction and Audio-Visual conditions across all measured emotion dimensions. A positive bar indicates a higher emotional response under the Squeeze Interaction condition, whereas a negative bar implies a stronger response under the Audio-Visual condition. Asterisks (*) indicate statistically significant differences (paired t-test, p $<$ 0.05).}
    \label{fig:bar}
\end{figure*}

The results (Table~\ref{tab:ttest_results} and Figure~\ref{fig:bar}) revealed that Attentiveness (t(39) = 2.37, p = .0229), Serenity (t(39) = 2.25, p = .0303), and Surprise (t(39) = 2.46, p = .0185) were significantly higher in the Squeeze Interaction condition compared to the Audio-Visual condition. These findings suggest that squeeze interaction enhances participants' focus, promotes relaxation, and introduces a greater sense of novelty during the experience. 

Two additional dimensions, Self-Assurance (p = .0925) and Fatigue (p = .0744), exhibited marginal trends but did not reach statistical significance. The slight increase in Self-Assurance may be attributed to participants' growing familiarity with the system and interaction mechanisms over time, leading to a subjective boost in confidence and perceived control. In contrast, Fatigue tended to be slightly higher in the  condition, potentially due to the physical engagement required by the squeezing interaction. Although not statistically significant, this trend suggests that while squeeze-based interactions may enhance engagement, it might also introduce additional physical effort that could contribute to increased fatigue toward the end of the session. No significant differences were observed in the remaining dimensions, including Fear, Guilt, Hostility, Joviality, Sadness, and Shyness (p-value $>$ .1).

\begin{table*}[!htbp]
\centering
\caption{Paired t-test results of emotional dimensions between Squeeze Interaction and Audio-Visual conditions}
\label{tab:haptic_emotion_full}
\footnotesize 
\setlength{\tabcolsep}{4pt} 
\renewcommand{\arraystretch}{1.1} 
\begin{tabular}{lrrrcrr}
\toprule
\textbf{Emotion} & \textbf{Mean Diff} & \textbf{$t$} & \textbf{Cohen's d} & \textbf{95\% CI} & \textbf{Raw $p$-value} & \textbf{Storey's Q $p$-value} \\
\midrule
Afraid                & 0.000  & 0.000  & 0.000  & [-0.33,0.33] & 1.000\phantom{$^{**}$}       & 0.716\phantom{$^{**}$}       \\
Scared                & -0.070 & -1.000 & -0.160 & [-0.49,0.17] & 0.323\phantom{$^{**}$}       & 0.477\phantom{$^{**}$}       \\
Frightened            & 0.000  & 0.000  & 0.000  & [-0.33,0.33] & 1.000\phantom{$^{**}$}       & 0.716\phantom{$^{**}$}       \\
Nervous               & -0.070 & -0.720 & -0.110 & [-0.45,0.22] & 0.474\phantom{$^{**}$}       & 0.481\phantom{$^{**}$}       \\
Jittery               & 0.000  & 0.000  & 0.000  & [-0.33,0.33] & 1.000\phantom{$^{**}$}       & 0.716\phantom{$^{**}$}       \\
Shaky                 & -0.200 & -2.450 & -0.390 & [-0.73,-0.05] & 0.019$^{*}$\phantom{$^{*}$} & 0.142\phantom{$^{**}$}       \\
Angry                 & 0.050  & 0.810  & 0.130  & [-0.20,0.46] & 0.421\phantom{$^{**}$}       & 0.477\phantom{$^{**}$}       \\
Hostile               & -0.030 & -0.330 & -0.050 & [-0.38,0.28] & 0.743\phantom{$^{**}$}       & 0.628\phantom{$^{**}$}       \\
Irritable             & 0.030  & 0.440  & 0.070  & [-0.26,0.40] & 0.660\phantom{$^{**}$}       & 0.584\phantom{$^{**}$}       \\
Scornful              & 0.000  & 0.000  & 0.000  & [-0.33,0.33] & 1.000\phantom{$^{**}$}       & 0.716\phantom{$^{**}$}       \\
Disgusted             & -0.050 & -0.810 & -0.130 & [-0.46,0.20] & 0.421\phantom{$^{**}$}       & 0.477\phantom{$^{**}$}       \\
Loathing              & -0.120 & -1.400 & -0.220 & [-0.56,0.11] & 0.168\phantom{$^{**}$}       & 0.437\phantom{$^{**}$}       \\
Guilty                & -0.030 & -0.330 & -0.050 & [-0.38,0.28] & 0.743\phantom{$^{**}$}       & 0.628\phantom{$^{**}$}       \\
Ashamed               & 0.120  & 1.400  & 0.220  & [-0.11,0.56] & 0.168\phantom{$^{**}$}       & 0.437\phantom{$^{**}$}       \\
Blameworthy           & 0.000  & 0.000  & 0.000  & [-0.33,0.33] & 1.000\phantom{$^{**}$}       & 0.716\phantom{$^{**}$}       \\
Angry.at.self         & -0.070 & -1.000 & -0.160 & [-0.49,0.17] & 0.323\phantom{$^{**}$}       & 0.477\phantom{$^{**}$}       \\
Disgusted.with.self   & -0.030 & -0.440 & -0.070 & [-0.40,0.26] & 0.660\phantom{$^{**}$}       & 0.584\phantom{$^{**}$}       \\
Dissatified.with.self & -0.100 & -2.080 & -0.330 & [-0.67,0.01] & 0.044$^{*}$\phantom{$^{*}$} & 0.189\phantom{$^{**}$}       \\
Sad                   & -0.070 & -1.000 & -0.160 & [-0.49,0.17] & 0.323\phantom{$^{**}$}       & 0.477\phantom{$^{**}$}       \\
Blue                  & -0.100 & -0.890 & -0.140 & [-0.47,0.19] & 0.378\phantom{$^{**}$}       & 0.477\phantom{$^{**}$}       \\
Downhearted           & 0.030  & 0.240  & 0.040  & [-0.29,0.37] & 0.812\phantom{$^{**}$}       & 0.663\phantom{$^{**}$}       \\
Alone                 & -0.030 & -0.220 & -0.030 & [-0.36,0.30] & 0.830\phantom{$^{**}$}       & 0.664\phantom{$^{**}$}       \\
Lonely                & -0.250 & -2.240 & -0.350 & [-0.69,-0.01] & 0.031$^{*}$\phantom{$^{*}$} & 0.189\phantom{$^{**}$}       \\
Happy                 & 0.100  & 0.810  & 0.130  & [-0.20,0.46] & 0.421\phantom{$^{**}$}       & 0.477\phantom{$^{**}$}       \\
Joyful                & 0.350  & 2.160  & 0.340  & [0.00,0.68]  & 0.037$^{*}$\phantom{$^{*}$} & 0.189\phantom{$^{**}$}       \\
Delighted             & 0.030  & 0.160  & 0.030  & [-0.30,0.35] & 0.875\phantom{$^{**}$}       & 0.671\phantom{$^{**}$}       \\
Cheerful              & 0.070  & 0.500  & 0.080  & [-0.25,0.41] & 0.618\phantom{$^{**}$}       & 0.581\phantom{$^{**}$}       \\
Excited               & 0.150  & 1.030  & 0.160  & [-0.17,0.49] & 0.309\phantom{$^{**}$}       & 0.477\phantom{$^{**}$}       \\
Enthusiastic          & 0.100  & 0.610  & 0.100  & [-0.23,0.43] & 0.544\phantom{$^{**}$}       & 0.524\phantom{$^{**}$}       \\
Lively                & 0.120  & 0.930  & 0.150  & [-0.18,0.48] & 0.360\phantom{$^{**}$}       & 0.477\phantom{$^{**}$}       \\
Energetic             & 0.250  & 1.880  & 0.300  & [-0.04,0.63] & 0.067\phantom{$^{**}$}       & 0.230\phantom{$^{**}$}       \\
Proud                 & 0.200  & 1.140  & 0.180  & [-0.15,0.51] & 0.263\phantom{$^{**}$}       & 0.477\phantom{$^{**}$}       \\
Strong                & 0.170  & 1.100  & 0.170  & [-0.16,0.51] & 0.280\phantom{$^{**}$}       & 0.477\phantom{$^{**}$}       \\
Confident             & 0.000  & 0.000  & 0.000  & [-0.33,0.33] & 1.000\phantom{$^{**}$}       & 0.716\phantom{$^{**}$}       \\
Bold                  & 0.070  & 1.140  & 0.180  & [-0.15,0.51] & 0.262\phantom{$^{**}$}       & 0.477\phantom{$^{**}$}       \\
Daring                & 0.220  & 1.140  & 0.180  & [-0.15,0.51] & 0.262\phantom{$^{**}$}       & 0.477\phantom{$^{**}$}       \\
Fearless              & 0.100  & 0.750  & 0.120  & [-0.21,0.45] & 0.457\phantom{$^{**}$}       & 0.477\phantom{$^{**}$}       \\
Alert                 & 0.200  & 1.350  & 0.210  & [-0.12,0.55] & 0.186\phantom{$^{**}$}       & 0.437\phantom{$^{**}$}       \\
Attentive             & 0.520  & 2.680  & 0.420  & [0.08,0.77]  & 0.011$^{*}$\phantom{$^{*}$} & 0.101\phantom{$^{**}$}       \\
Concentrating         & 0.120  & 0.870  & 0.140  & [-0.19,0.47] & 0.391\phantom{$^{**}$}       & 0.477\phantom{$^{**}$}       \\
Determined            & 0.150  & 1.000  & 0.160  & [-0.17,0.49] & 0.323\phantom{$^{**}$}       & 0.477\phantom{$^{**}$}       \\
Shy                   & -0.070 & -1.360 & -0.210 & [-0.55,0.12] & 0.183\phantom{$^{**}$}       & 0.437\phantom{$^{**}$}       \\
Bashful               & -0.100 & -0.810 & -0.130 & [-0.46,0.20] & 0.421\phantom{$^{**}$}       & 0.477\phantom{$^{**}$}       \\
Sheepish              & 0.050  & 0.700  & 0.110  & [-0.22,0.44] & 0.486\phantom{$^{**}$}       & 0.481\phantom{$^{**}$}       \\
Timid                 & -0.070 & -1.360 & -0.210 & [-0.55,0.12] & 0.183\phantom{$^{**}$}       & 0.437\phantom{$^{**}$}       \\
Sleepy                & -0.030 & -0.170 & -0.030 & [-0.36,0.30] & 0.864\phantom{$^{**}$}       & 0.671\phantom{$^{**}$}       \\
Tired                 & -0.100 & -0.810 & -0.130 & [-0.46,0.20] & 0.421\phantom{$^{**}$}       & 0.477\phantom{$^{**}$}       \\
Sluggish              & -0.400 & -3.250 & -0.510 & [-0.87,-0.16] & 0.002$^{**}$& 0.049$^{*}$\phantom{$^{*}$}       \\
Drowsy                & -0.200 & -1.210 & -0.190 & [-0.52,0.14] & 0.232\phantom{$^{**}$}       & 0.477\phantom{$^{**}$}       \\
Calm                  & 0.100  & 0.430  & 0.070  & [-0.26,0.40] & 0.668\phantom{$^{**}$}       & 0.584\phantom{$^{**}$}       \\
Relaxed               & 0.300  & 2.020  & 0.320  & [-0.02,0.66] & 0.050\phantom{$^{**}$}       & 0.189\phantom{$^{**}$}       \\
At.ease               & 0.480  & 3.220  & 0.510  & [0.16,0.86]  & 0.003$^{**}$& 0.049$^{*}$\phantom{$^{*}$}       \\
Amazed                & 0.320  & 2.060  & 0.330  & [-0.01,0.66] & 0.046$^{*}$\phantom{$^{*}$} & 0.189\phantom{$^{**}$}       \\
Surprised             & 0.120  & 1.000  & 0.160  & [-0.17,0.49] & 0.323\phantom{$^{**}$}       & 0.477\phantom{$^{**}$}       \\
Astonished            & 0.320  & 2.690  & 0.430  & [0.08,0.77]  & 0.010$^{*}$\phantom{$^{*}$} & 0.101\phantom{$^{**}$}       \\
\bottomrule
\multicolumn{7}{l}{\footnotesize{$^{*}p < .05$, $^{**}p < .01$. CI = Confidence Interval}} \\
\end{tabular}
\vspace{-1.0em}
\end{table*}

To enhance the interpretability of the results, we calculated the mean difference ($\Delta$), 95\% confidence interval (95\% CI), raw $p$-value, $p$-values corrected using Storey's q-value method, and paired Cohen's $d$ with effect size (including 95\% CI) for each dimension, aiming to facilitate understanding of the practical magnitude of the effects.

The results (Table~\ref{tab:haptic_emotion_full}) showed that the At Ease subscale (t(39) = 3.22, raw $p$-value = 0.0026, Storey's q-value corrected $p$-value = 0.0487, mean difference ($\Delta$) = 0.48, 95\% CI = [0.18, 0.77], Cohen's $d$= 0.51, effect size (95\% CI) = [0.16, 0.86]) under the Serenity domain of Squeeze Interaction was significantly superior to the Audio-Visual condition. Under the same Serenity domain, the Relaxed subscale (t(39) = 2.02, raw $p$-value = 0.0503, Storey's q-value corrected $p$-value = 0.1889, mean difference ($\Delta$) = 0.30, 95\% CI = [0.00, 0.60], Cohen's $d$= 0.32, effect size (95\% CI) = [-0.02, 0.66]) and the Amazed subscale (t(39) = 2.06, raw $p$-value = 0.0460, Storey's q-value corrected $p$-value = 0.1889, mean difference ($\Delta$) = 0.32, 95\% CI = [0.01, 0.64], Cohen's $d$= 0.33, effect size (95\% CI) = [-0.01, 0.66]) under the Surprise domain did not reach statistical significance after correction. However, their raw $p$-values, mean differences, 95\% CIs, and Cohen's $d$ values all approached the significance threshold, with effect directions consistent with the research hypotheses. This suggests potential underlying effects that may not have been fully captured due to limited measurement precision. Similarly, the Attentive subscale (t(39) = 2.68, raw $p$-value = 0.0108, Storey's q-value corrected $p$-value = 0.1014, mean difference ($\Delta$) = 0.52, 95\% CI = [0.13, 0.92], Cohen's $d$= 0.42, effect size (95\% CI) = [0.08, 0.77]) under the Attentiveness domain and the Astonished subscale  (t(39) = 2.69, raw $p$-value = 0.0105, Storey's q-value corrected $p$-value = 0.1014, mean difference ($\Delta$) = 0.32, 95\% CI = [0.08, 0.57], Cohen's $d$= 0.43, effect size (95\% CI) = [0.08, 0.77]) under the Surprise domain of Squeeze Interaction were significantly superior to the Audio-Visual condition in uncorrected tests. Although their significance weakened after correction, the mean differences, 95\% CIs, and effect sizes clearly demonstrate that the advantages of the Squeeze Interaction condition over the Audio-Visual condition in these two subscales are real and observable, with a high likelihood of achieving corrected significance with further validation. In addition, the Joyful subscale (t(39) = 2.16, raw $p$-value = 0.0373, Storey's q-value corrected $p$-value = 0.1889, mean difference ($\Delta$) = 0.35, 95\% CI [0.02, 0.68], Cohen's $d$= 0.34, effect size (95\% CI) = [0.00, 0.68]) under the Joviality domain exhibited a non-negligible marginal trend.

To verify the impact of counterbalancing the experimental order (Participants P1–20 first experienced the Squeeze Interaction condition followed by the Audio-Visual condition; Participants P21–40 first experienced the Audio-Visual condition followed by the Squeeze Interaction condition) on the condition effect (Squeeze Interaction vs. Audio-Visual experience), a mixed-design analysis of variance (ANOVA) was performed on the eleven emotional domains. The independent variables were experimental order (Squeeze Interaction-first vs. Squeeze Interaction-second) and experimental condition (Squeeze Interaction vs. Audio-Visual), and Bonferroni correction was applied to the results to control for false positives.

Regarding the main effect of experimental order, all emotional domains exhibited negligible effect sizes ($\eta_p^2$): Fear ($\eta_p^2$ = 0.0119, corrected $p$-value = 0.3534), Hostility ($\eta_p^2$ = 0.0049, corrected $p$-value = 0.6664), Guilt ($\eta_p^2$ = 0.0313, corrected $p$-value = 0.2747), Sadness ($\eta_p^2$ = 0.0084, corrected $p$-value = 0.5732), Joviality ($\eta_p^2$ = 0.0022, corrected $p$-value = 0.7758), Self-Assurance ($\eta_p^2$ = 0.0227, corrected $p$-value = 0.3534), Attentiveness ($\eta_p^2$ = 0.0058, corrected $p$-value = 0.6405), Shyness ($\eta_p^2$ = 0.0027, corrected $p$-value = 0.7505), Fatigue ($\eta_p^2$ = 0.0001, corrected $p$-value = 0.9461), Serenity ($\eta_p^2$ = 0.0019, corrected $p$-value = 0.7909), and Surprise ($\eta_p^2$ = 0.0140, corrected $p$-value = 0.4668). None of the corrected $p$-values reached statistical significance (Table~\ref{tab:anova_merge_font_fix}). Furthermore, while the mean squared error (MSE) values for some emotions were relatively large, all F-statistics were extremely small. This indicates that the effect induced by experimental order was far smaller than the random variability in the data.

\begin{table}[!htbp]
\centering
\caption{ANOVA results for emotion (Order and Order:Condition effects)}
\label{tab:anova_merge_font_fix}
\footnotesize 
\setlength{\tabcolsep}{4pt} 
\renewcommand{\arraystretch}{1.1} 
\begin{tabular}{lrlrrrl}
\toprule[1pt] 
\textbf{Emotion} & & \textbf{Effect} & \textbf{MSE} & \textbf{F} & \textbf{pes} & \textbf{Pr(>F)} \\
\midrule[0.2pt] 
Fear       & & Order               & 0.246 & 0.457 & 0.012 & 0.503 \\
 & & Order:Condition     & 0.022 & 7.538 & 0.166 & 0.009** \\
Hostility  & & Order               & 0.222 & 0.189 & 0.005 & 0.666 \\
 & & Order:Condition     & 0.034 & 4.488 & 0.106 & 0.041* \\
Guilt      & & Order               & 0.547 & 1.228 & 0.031 & 0.275 \\
& & Order:Condition     & 0.013 & 5.074 & 0.118 & 0.030* \\
Sadness    & & Order               & 0.559 & 0.323 & 0.008 & 0.573 \\
& & Order:Condition     & 0.061 & 5.994 & 0.136 & 0.019* \\
Joviality  & & Order               & 1.731 & 0.082 & 0.002 & 0.776 \\
& & Order:Condition     & 0.200 & 1.064 & 0.027 & 0.309 \\
Self-Assurance & & Order           & 1.370 & 0.883 & 0.023 & 0.353 \\
& & Order:Condition & 0.113 & 0.691 & 0.018 & 0.411 \\
Attentiveness  & & Order           & 0.903 & 0.222 & 0.006 & 0.641 \\
& & Order:Condition & 0.228 & 0.055 & 0.001 & 0.816 \\
Shyness    & & Order               & 0.274 & 0.103 & 0.003 & 0.751 \\
& & Order:Condition     & 0.030 & 8.479 & 0.182 & 0.006** \\
Fatigue    & & Order               & 1.517 & 0.005 & 0.000 & 0.946 \\
& & Order:Condition     & 0.197 & 0.670 & 0.017 & 0.418 \\
Serenity   & & Order               & 0.954 & 0.071 & 0.002 & 0.791 \\
& & Order:Condition     & 0.339 & 0.692 & 0.018 & 0.411 \\
Surprise   & & Order               & 1.134 & 0.540 & 0.014 & 0.467 \\
& & Order:Condition     & 0.196 & 5.965 & 0.136 & 0.019* \\
\bottomrule
\multicolumn{6}{l}{$^*$Significant at $p < .05$, $p < .01$}

\end{tabular}
\vspace{-1.0em}
\end{table}

However, regarding the interaction effect between experimental order and experimental condition, Fear (corrected $p$-value = 0.0092), Hostility (corrected $p$-value = 0.0407), Guilt (corrected $p$-value = 0.0302), Sadness (corrected $p$-value = 0.0191), Shyness (corrected $p$-value = 0.0060), and Surprise (corrected $p$-value = 0.0194) exhibited a certain level of statistical significance (Table~\ref{tab:anova_merge_font_fix}). For the four emotion domains (Fear, Guilt, Hostility, Sadness), the key difference was that the magnitude of change was significantly smaller in the squeeze interaction-first then Audio-Visual order than in the Audio-Visual-first then squeeze interaction order. 
This is highly interpretable as an asymmetrical carryover effect~\cite{poulton1973unwanted,senn2002cross}. Specifically, the Squeeze Interaction was sufficiently potent to alleviate negative affect to near-basal levels. Consequently, participants in the Squeeze-First group experienced a floor effect~\cite{liu2021t} during the subsequent Audio-Visual phase, as there was little remaining negative affect to reduce. This structural constraint of the crossover design~\cite{senn2002cross}, indirectly validates the superior efficacy of the squeeze interaction condition. Additionally, participants reported significantly higher Surprise scores(Figure~\ref{fig:barall}) when they first experienced the Squeeze Interaction condition (compared to the Audio-Visual-first order). This may be attributed to the plush toy's inclusion being a fresher, less common experience in the VR environment for the participants.


In summary, the variations induced by the interaction between order and condition are logically sound and interpretable. Far from invalidating the results, these order-dependent patterns, and the cross-sectional comparisons of the first experimental stage, actually reinforce our findings regarding the effectiveness of the squeeze interaction.

We also conducted normality tests on the within-subjects difference scores (Squeeze Interaction condition scores minus Audio-Visual condition scores) across emotional subscales (see figure for more details in Supplemental Material). Visual inspection of the normality histograms revealed that the majority of subscales generally followed or exhibited a trend of approximate normality, but failed to fully conform to normality, potentially due to limitations in data measurement precision. Subscales that deviated significantly from normality were almost exclusively categorized which are not positive enough, with anomalies predominantly concentrated in the negative difference score range. These subscales included Afraid, Angry, Angry at Self, Blameworthy, Bold, Disgusted, Disgusted with Self, Dissatisfied with Self, Frightened, Guilty, Irritable, Jittery, Sad, Scared, Scornful, Shaky, Sleepy, Shy, and Timid. This phenomenon is likely attributed to the fact that negative emotion scores in the Squeeze Interaction condition were significantly lower than those in the Audio-Visual condition for the vast majority of participants.

These results collectively indicate that while squeeze interaction does not uniformly influence all emotional states, it significantly enhances engagement-related and calming emotional responses in immersive environments.

\vspace{-0.2cm}

\subsubsection{\textbf{Subjective Experience Evaluation Questionnaire: Results of Sixteen Universal Questions}}
We conducted a detailed analysis of subjective evaluation responses collected after participants experienced the VR Calm Plus application. The first analysis/evaluation included 16 universal questions covering immersion, emotional regulation, scene design, interaction clarity, and overall user satisfaction.

The demographic characteristics (Table~\ref{table:demograph_participants}) indicates that 57.5\% of the participants had VR experience at the ``never'' (no prior exposure) or ``initial exposure'' (one-time or very limited exposure) level, 62.5\% of them had meditation experience at the ``never'' or ``initial exposure'' level, and 40\% of them had both experiences at the ``never'' or ``initial exposure'' level.

\begin{figure*}[htbp] 
    \centering
    \includegraphics[width=0.8\linewidth]{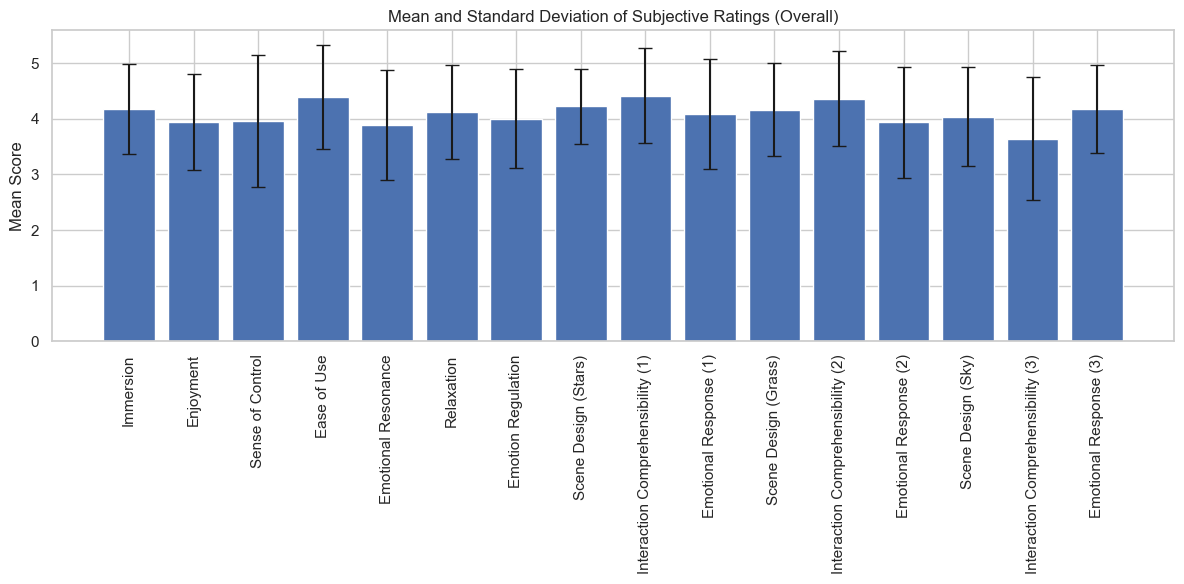}
    \Description{Most mean scores fall between 3 and 5, indicating participants generally gave favorable evaluations for aspects like immersion, enjoyment, ease of use, emotional response (2), relaxation, and emotion regulation.}
    \caption{Overall Average Scores and Standard Deviation of Subjective Questionnaire. The Y-axis represents the score, and the X-axis denotes the abbreviated question, with the text in parentheses referring to a specific scenario.}
    \label{fig:subjectiveQuestionnaire-16all}
\end{figure*}

The results of descriptive statistics (Figure~\ref{fig:subjectiveQuestionnaire-16all}) revealed consistently high ratings across both conditions. This indicates that the design of VR Calm Plus was generally well accepted by the participants. Participants reported particularly strong agreement on items such as ``\textit{Ease of Use}'' (M = 4.65, SD = 0.48 in the Squeeze Interaction condition) and ``\textit{Scene Design (Stars)}'' (M = 4.50, SD = 0.68), indicating a generally positive perception of the system's usability and aesthetic composition. The internal consistency of the scale was excellent (Cronbach's $\alpha = .92$), validating the reliability of the instrument.

To evaluate the influence of squeeze-based haptic feedback, we compared responses between the Squeeze Interaction and Audio-Visual conditions (see Figure S11 in Supplemental Material). Independent samples t-tests (Table~\ref{tab:subjectivettest_results}) showed that the Squeeze Interaction condition significantly outperformed the Audio-Visual condition on multiple key dimensions. These results suggest that incorporating squeeze-based haptic feedback meaningfully enhances users' sense of control, emotional engagement, and perceived effectiveness of emotion regulation. 

\begin{table*}
\centering
\Description{Squeeze Interaction outperforms Audio-Visual in several dimensions: Enjoyment (Squeeze Interaction: 4.15 vs. Audio-Visual: 3.73), Sense of Control (4.40 vs. 3.52), Ease of Use (4.65 vs. 3.12), Interaction Comprehensibility (1) (4.65 vs. 4.17), and Emotional Response (1) (4.42 vs. 3.75).}
\caption{Detailed Group Comparison of Subjective Ratings Between Squeeze Interaction and Audio-Visual Conditions}
\label{tab:subjectivettest_results}
\scalebox{0.8}{
\begin{tabular}{lccccc}
\toprule
\textbf{Question} & \textbf{Squeeze Interaction} & \textbf{Squeeze Interaction} & \textbf{Audio-Visual} & \textbf{Audio-Visual} & \textbf{p-value} \\
&\textbf{Mean}&\textbf{SD}&\textbf{Mean}&\textbf{SD}&\\
\midrule
Immersion & 4.30 & 0.76 & 4.05 & 0.85 & 0.168 \\
Enjoyment & 4.15 & 0.77 & 3.73 & 0.91 & 0.027*  \\
Sense of Control & 4.40 & 0.81 & 3.52 & 1.34 & 0.001*  \\
Ease of Use & 4.65 & 0.48 & 4.12 & 1.18 & 0.012* \\
Emotional Resonance & 4.12 & 0.85 & 3.65 & 1.05 & 0.030*  \\
Relaxation & 4.25 & 0.81 & 4.00 & 0.88 & 0.189 \\
Emotion Regulation & 4.17 & 0.87 & 3.83 & 0.87 & 0.077  \\
Scene Design (Stars) & 4.25 & 0.67 & 4.20 & 0.69 & 0.743  \\
Interaction Comprehensibility (1) & 4.65 & 0.53 & 4.17 & 1.03 & 0.012*  \\
Emotional Response (1) & 4.42 & 0.59 & 3.75 & 1.17 & 0.002*  \\
Scene Design (Grass) & 4.15 & 0.83 & 4.17 & 0.84 & 0.894  \\
Interaction Comprehensibility (2) & 4.55 & 0.60 & 4.17 & 1.03 & 0.052  \\
Emotional Response (2) & 4.10 & 0.93 & 3.77 & 1.05 & 0.146  \\
Scene Design (Sky) & 4.05 & 0.88 & 4.03 & 0.92 & 0.901 \\
Interaction Comprehensibility (3) & 3.60 & 1.13 & 3.67 & 1.10 & 0.764  \\
Emotional Response (3) & 4.22 & 0.80 & 4.12 & 0.79 & 0.576  \\
\bottomrule
\multicolumn{6}{l}{\footnotesize{$^{*}$Significant at $p < .05$}}
\end{tabular}
}
\end{table*}

These findings support the hypothesis that squeeze feedback enhances emotional engagement and experiential depth in VR-based emotion regulation applications.

\subsubsection{\textbf{Results of physiological data}}
\begin{figure*}[htbp]
    \centering
    \includegraphics[width=0.9\linewidth]{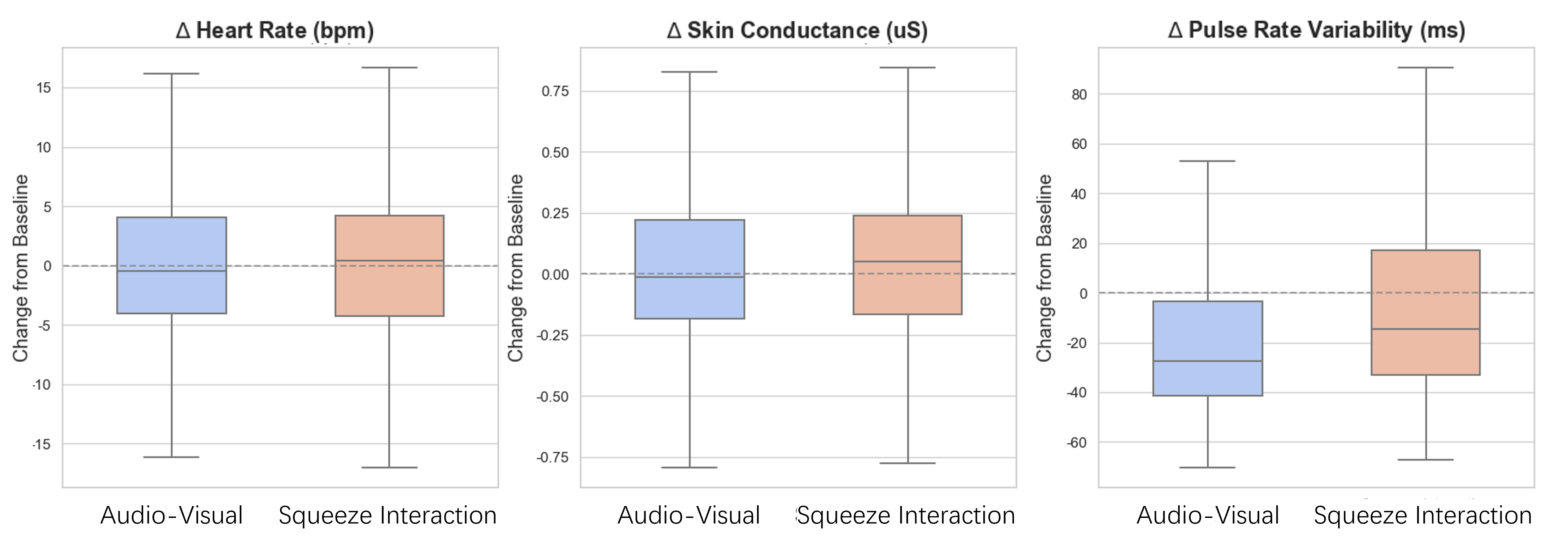}
    \Description{Three side-by-side boxplots comparing Audio-Visual and Squeeze Interaction conditions. 
    (1) Heart Rate Delta: Squeeze Interaction shows a lower median value (more negative) than Audio-Visual, indicating deeper relaxation. 
    (2) Skin Conductance Delta: Squeeze Interaction shows a slightly positive median, higher than Audio-Visual, indicating maintained engagement. 
    (3) PRV Delta: Audio-Visual shows a significant drop (negative value), while Squeeze Interaction remains close to zero (baseline), indicating preserved autonomic balance.}
    \caption{Comparison of physiological changes from baseline ($\Delta$) between Audio-Visual and Squeeze Interaction conditions. The dashed line at $y=0$ represents the individual baseline level. Results indicate that Squeeze Interaction induced significantly deeper relaxation (lower $\Delta$HR), sustained emotional engagement (higher $\Delta$GSR), and preserved autonomic flexibility (stable $\Delta$PRV) compared to the Audio-Visual condition.}
    \label{fig:bioData}
\end{figure*}

We analyzed the change from baseline ($\Delta$) for three physiological metrics, Heart Rate ($\Delta$HR), Galvanic Skin Response ($\Delta$GSR), and Pulse Rate Variability ($\Delta$PRV), to determine the objective impact of squeeze interactions. Table~\ref{tab:physio_results} and Figure~\ref{fig:bioData} summarize the statistical comparisons between the Squeeze Interaction and Audio-Visual conditions.

\begin{table*}[htbp]
  \centering
  \caption{Comparison of Physiological Changes from Baseline ($\Delta$) between Conditions.}
  \label{tab:physio_results}
  \begin{tabular}{lcccc}
    \toprule
    \textbf{Metric} & \textbf{Squeeze Interaction} & \textbf{Audio-Visual} & \textbf{Mean Diff. (S - V)} & \textbf{\textit{p}-value} \\
    & (Mean $\Delta$) & (Mean $\Delta$) & & \\
    \midrule
    Heart Rate (bpm) & -0.54 & -0.13 & -0.41 & $< .001^{***}$ \\
    Skin Conductance ($\mu$S) & 0.08 & 0.02 & 0.05 & $< .001^{***}$ \\
    Pulse Rate Variability (ms) & -4.61 & -20.42 & 15.81 & $< .001^{***}$ \\
    \bottomrule
    \multicolumn{5}{l}{\footnotesize Note: $\Delta$ represents the deviation from the individual baseline. $^{***}p < .001$.}
  \end{tabular}
\end{table*}

\paragraph{Heart Rate ($\Delta$HR).}
The analysis revealed a statistically significant difference in heart rate reduction between conditions ($p < .001$). While both conditions induced a decrease relative to baseline, the Squeeze Interaction elicited a significantly deeper reduction ($\Delta M = -0.54$) compared to the Audio-Visual condition ($\Delta M = -0.13$). This indicates that the embodied act of squeezing the plush toy facilitates a superior state of physiological relaxation compared to the purely visual experience.

\paragraph{Skin Conductance ($\Delta$GSR).} 
A significant difference was also observed in electrodermal activity ($p < .001$). The Squeeze Interaction resulted in a significantly higher increase in skin conductance ($\Delta M = 0.08$) compared to the Audio-Visual condition ($\Delta M = 0.02$). Although the absolute difference appears small, the statistical significance suggests a robust effect: despite being physically relaxed (lower HR), participants remained emotionally present. The squeeze interaction effectively mitigates the physiological disengagement often associated with passive viewing.

\paragraph{Pulse Rate Variability ($\Delta$PRV).} 
The most statistically significant divergence emerged in autonomic regulation ($p < .001$). The Audio-Visual condition led to a sharp, significant decline in variability ($\Delta M = -20.42$), indicating physiological rigidity or fatigue. In contrast, the Squeeze Interaction significantly preserved PRV levels near the baseline ($\Delta M = -4.61$). This preservation suggests that the rhythmic squeezing acts as a physiological buffer, maintaining the body's autonomic flexibility.

\paragraph{Summary.}
Collectively, these physiological signatures characterize a state of ``active relaxation'': the Squeeze Interaction significantly reduces cardiac workload (lower HR) while simultaneously maintaining autonomic flexibility (stable PRV) and emotional engagement (higher GSR). These objective findings corroborate the subjective reports, confirming that embodied squeeze interaction transforms the VR experience from passive consumption to an emotionally regulated engagement.

\subsection{Qualitative Findings}
\subsubsection{The System Design and User Strategies of the ``VR Calm Plus'' Influence Users' Emotional Engagement}

The level of perceived emotional engagement in VR emotional experiences is closely related to users' attention focus. Most user reports indicate that a stronger sense of immersion was perceived in the Squeeze Interaction condition than in the Audio-Visual condition, with users demonstrating higher levels of focus and engagement.

\textit{1) Enhancement of Focus.} Participants frequently reported improvements in focus. The vast majority of participants (P1, P4, P7, P10, P11, P13-P20, P24, P26-P30, P32, P34, P36, P38, P39) explicitly stated that the squeezing interaction enhanced their focus. The core reasons for this included the gamification appeal of rhythmic pressing in the second stage and the color changes in visual feedback in the first stage. Additionally, the timely consistency of the coordination among squeeze-based haptic, visual, and auditory feedback also contributed to this outcome, as it redirected participants' attention to elements that aligned with emotional resonance.
In contrast, a subset of participants (P4, P6, P8, P13, P14, P16, P21, P24, P27, P35) indicated that without the involvement of physical haptic feedback, they were prone to being distracted by disorganized thoughts, which ultimately led to reduced focus.

\textit{2) Enhancement of Emotion Engagement.} Squeeze-based haptic feedback could further deepen participants' focus. A number of participants (P2, P10, P14, P15, P20, P23, P38) reported that the pressing interaction had achieved effective depth of experience, and they perceived that their engagement was significantly enhanced 

\textit{3) Enhancement of Immersion Experience.} The majority of participants (P4, P7, P11, P13, P15-P17, P23, P24, P26, P32, P35, P38) reported that the pressing interaction enhanced their sense of immersion. Meanwhile, a subset of users indicated that the music rhythm-following task in the second stage had a particularly prominent effect on improving immersion; in the engaging ``rhythm game,'' some even entered a highly focused flow state.
\begin{quote}
    \textit{``It gives me the kind of feeling that I'm entering a flow state.''} (P7).
\end{quote}

\textit{4) Enhancement of Sense of Control.} The sense of control is crucial for relaxation in high stress states. Participant P32 stated, \begin{quote}\textit{``If you want to vent your negative emotions, the first one with a strong sense of control is good.''} (P32).\end{quote} Many participants mentioned that the squeeze interaction significantly enhanced their sense of control in the VR environment (P1, P3-7, P9, P15, P18-P20, P23, P31), and all held positive opinions about this—they believed that the enhanced sense of control effectively improved their emotional experience, focus, and engagement (P15, P19). Additionally, some participants were still not satisfied with the level of sense of control in the Squeeze Interaction condition, stating that adding more controllable elements could further enhance the sense of control for a better experience (P23).

The sense of control has an extremely critical impact on participants' emotions.

On one hand, the generation of the sense of control is closely related to squeeze interaction. Participants' reports indicated that the sense of control mainly originates from the interaction process of `squeezing force - visual feedback.' Some participants explained that two aspects, the visualization of squeeze force and the timely responsive changes in particle effects and colors upon squeezing”, are the main reasons for the enhanced sense of control (P5, P9, P20). Enhanced sense of control is often accompanied by a high arousal level (P32), which makes a significant contribution to the maintenance and recovery of high-arousal positive emotions such as ``excited,'' and ``enthusiastic,'' as well as to indicators like ``happy,'' ``lively,'' ``cheerful,'' and ``energetic.''

On the other hand, both excessively low and excessively high sense of control pose risks. Excessively low or even lost sense of control may trigger negative emotions: Participant P18 stated that they felt worried about the particle explosion effect at the end of the first stage, describing it as \begin{quote}\textit{``like being squeezed and burst.''} (P18).\end{quote} Participant P39 had a negative experience due to mandatory breathing guidance in the third stage. Meanwhile, excessively high sense of control may lead to the possibility of shifting the focus of experience from regulating one's own emotions to controlling the external environment (P32, P36).

The effect of the sense of control in emotion regulation cannot be ignored, but its threshold still needs to be properly managed. Therefore, it is crucial to integrate multimodal channels and introduce an emotion adaptive system for users. This system can properly adjust the intensity, mode, or rhythm of squeeze responses to support personalized emotional needs.

\textit{5) Enhancement of Sense of Engagement.} Within a reasonable range of sense of control, a subset of participants reported having an engaging and pleasant experience in the Squeeze Interaction condition (P1, P5, P9, P25, P29, P31, P33, P36, P40), while they gave negative evaluations of lacking engagement in the Audio-Visual condition (P19, P20).

\subsubsection{The System Design and User Strategies of the ``VR Calm Plus'' Influence Users' Calm Emotional State}

In VR emotional experiences, it is well recognized that music and scenes exert undeniable effects on emotional relaxation and calm emotional states. Beyond this, squeeze interaction has also demonstrated significant potential.

\textit{1) Sense of Connection.} The sense of connection with the VR environment reflects the degree of interaction, resonance, and perception. It is largely influenced by multiple factors such as focus, engagement, sense of immersion, sense of control, and sense of engagement, and is centrally embodied in the coherence across multiple sensory channels. Participant P1 reported that physical squeeze-based haptic feedback, as a medium channel, significantly enhanced their sense of connection with the VR visual and auditory environment. Participant P8 argued that the squeeze-based haptic channel should be further enhanced by incorporating more feedback methods such as vibration. Similarly, many reports indicated that haptic contact materials have a significant impact on the factors of experience, including personal preferences, uneven surfaces, and internal structures affect the pressing feel and force (P11, P17, P25, P27, P30, P31, P36, P37, P40). And emphasized the compatibility of sensory mapping between the haptic feel of the plush material and particle sphere visual effects (P37). Participant P36 also proposed adding ``a warmth of sunlight'' temperature sensory channel to enrich the modalities.

\textit{2) Squeezing Actions.} \begin{quote}\textit{``When I squeezed it, the color changed, and this change was very regular, which made me feel really good.''} (P14).\end{quote}
\begin{quote}\textit{``The second stage allowed me to squeeze along with the rhythm of the music, which made me feel quite comfortable and relaxed.''} (P17).\end{quote}
Furthermore, the action of squeezing a plush toy or even objects made of other materials itself was regarded by many participants as a natural and effective way to relieve stress (P3-P5, P7, P15, P17, P20, P21, P23, P26, P34). However, the physical exertion caused by prolonged and high frequency squeezing, as well as the bodily movement itself, should not be overlooked. It may lead to a certain degree of fatigue, and some reports indicated that this fatigue has caused a certain level of negative impact (P4, P7, P8, P11-P13, P20, P24, P25, P28, P29, P32, P37).

\textit{3) Cognitive Resources.} Given that the background music in the second stage featured an upbeat and lively rhythm, engaging in interaction in time with this rhythm required the consumption of a certain amount of cognitive resources (P4, P5, P9-P11, P16, P20, P23, P31, P32, P37, P38). Specifically, participants who attempted to align their interactions with every musical note reported experiencing a certain level of cognitive load and fatigue, which undermined their emotional experience but did not result in negative effects. As one participant noted, \begin{quote}``\textit{It didn't make me uncomfortable, nor did it trigger negative emotions.}'' (P32).\end{quote} In contrast, participants who only attempted to follow the key beats reported an increase in emotional arousal and perceived positive emotions such as ``excited'' and ``enthusiastic''.

Overall, participants generally agreed that high-intensity squeeze interaction was more conducive to meeting the need for emotional stress release. However, if the goal was to achieve emotional calmness, they expressed a stronger preference for squeeze interaction that they could follow with ease, like participant P10 stated, \begin{quote}\textit{``There were definitely no negative effects. If it were replaced with one I could keep up with, the experience would be better.}'' (P10).\end{quote} Another favorable option was the fully user-driven squeeze interaction mode in the first stage, as reflected by P5, \begin{quote}``\textit{The first stage wasn't a burden because I could rely on my own rhythm.}'' (P5).\end{quote}

\vspace{-0.5em}
\section{Discussion}

\subsection{Squeeze-based Haptic Feedback as a Mechanism for Emotional Amplification in VR}

Our findings suggest that incorporating squeeze feedback into immersive experiences meaningfully enhances emotional engagement—particularly by increasing attentiveness, serenity, and a sense of novelty. These results align with theories of embodied cognition, which emphasize that bodily actions and sensorimotor experiences directly shape emotional and cognitive states~\cite{Foglia2013}. The haptic act of squeezing, as implemented in VR Calm Plus, offers a form of embodied emotional modulation that reinforces affective feedback from the environment.

Moreover, the observed emotional shifts support the potential of squeeze-based haptic feedback as an emotional amplifier in VR. For example, the decrease in ``joyful'', ``excited'', and ``enthusiastic'' scores, especially in the Audio-Visual condition, may be attributed to novelty effects and emotional adaptation. Participants may have initially anticipated an engaging experience, resulting in higher baseline scores for these high-arousal emotions. As the session progressed, the lack of multisensory feedback in the Audio-Visual condition could have led to emotional habituation and reduced excitement. In contrast, squeeze-based haptic feedback appeared to help maintain or restore these positive emotions, likely due to the added embodied interaction. Specifically, as reported by participants, the squeezing action input and particle scaling feedback that align with embodied cognition have effectively enhanced their sense of immersion, which fosters playfulness and engagement. Prior work in affective computing has shown that squeeze interactions can deepen affective experiences~\cite{7320966}, but their role in emotional transformation, rather than merely expression. By providing immediate and intuitive physical feedback, squeeze-based haptic modalities may facilitate more immersive and active emotion regulation strategies.

\subsection{Designing for Multisensory Coherence and Interaction Transparency}
Participants in the Squeeze Interaction condition reported greater clarity in interaction understanding, emotional resonance, and perceived control. This highlights the importance of multisensory coherence, the alignment of touch, sight, and sound, in crafting emotionally effective VR experiences. According to media richness theory~\cite{daft1986media}, richer modalities provide more immediate feedback and reduce ambiguity in communication. Our results extend this perspective into emotion-centered design, suggesting that squeeze-based haptic channels enhance the clarity of emotional cues, making interactions more legible and emotionally salient (it seems a way to process the interaction and signal transmission directly).

From a design standpoint, this suggests that squeeze augmentation can enhance the transparency of affective systems, particularly in scenarios where emotional intent must be intuitively communicated. In the future, this fast, direct, and transparent mode of emotional transmission through squeeze interaction can be further extended to various VR environments. For example, users' squeeze-based haptic feedback can be used to complement passive emotion detection methods, where the goal is not merely to simulate emotion, but to scaffold users' awareness and regulation of their own affective trajectories in immersive environments~\cite{colombo2021virtual}. Furthermore, some participants, in pursuit of higher sensory consistency, took the initiative to bear the cognitive load caused by high-intensity interaction during the second phase. However, the effects of their emotional experience were instead undermined. In contrast, the first phase, which offered a higher degree of freedom, maintained relatively stable effects of the emotional experience. This phenomenon further indicates that VR emotional experience not only relies on the multisensory coherence but also requires avoiding the interference of cognitive load on the clarity of emotional transmission. Designers should therefore carefully balance sensory integration with interactional simplicity, ensuring that coherence does not unintentionally escalate into cognitive strain. This aligns with prior findings in cognitive load theory~\cite{sweller1988cognitive,sweller2011cognitive}, suggesting that overstimulation may detract from the affective benefits of immersive media.

\subsection{Toward Bi-Directional Emotion-Squeeze-based Haptic Alignment Through Multimodal Integration}
Reflections on the experimental results led us to recognize that users' emotional needs are dynamic. For instance, when the goal of emotional regulation is mental calmness, high-intensity squeeze interaction feedback is not required and should even be avoided. As reported by users, the rhythm-following squeezing process in the second phase, which involved high-frequency squeeze interaction, facilitated the release of negative emotions and the arousal of positive emotions. In contrast, the first phase, though also incorporating squeeze interaction, exhibited a significant emotional soothing effect in reports, due to its completely free and unrestricted interaction rhythm. Therefore, consideration should be given to how to dynamically adapt to user needs. Based on the idea of adaptive feedback loops and the concept of bidirectional alignment~\cite{shen2024bidirectionalhumanaialignmentsystematic}, future work can advance toward bi-directional emotion-squeeze-based haptic alignment by integrating physiological sensing modalities (e.g., heart rate, galvanic skin response, respiration rate)~\cite{kyriakou2019detecting,yu2021vibreathe} with intelligent squeeze actuation. This approach would allow systems not only to deliver squeeze feedback but also to dynamically interpret users' internal emotional states and respond accordingly, forming a closed-loop regulation mechanism. Such affect-adaptive systems could personalize the intensity, pattern, or rhythm of squeeze responses to better align with users' current affective needs.

Additionally, participants proposed that enriching feedback from modalities other than visual and auditory ones would further enhance the sense of connection with the VR environment and better map multisensory coherence, such as environmental temperature and vibration feedback. This aligns with evidence that VR-based emotion regulation benefits from heightened multisensory engagement and presence~\cite{riches2021virtual,liszio2018relaxing,10.1145/3643834.3661570}, and that active engagement modalities can reduce stress and increase positive affect~\cite{10316434}.

Advances in hardware and material science offer promising directions for transforming the physical characteristics of squeeze-based haptic interfaces. For instance, incorporating automated damping mechanisms can regulate resistance and help balance pressure distribution during interactions, thereby improving comfort and reducing localized strain~\cite{chen2026deformable}. Similarly, using multi-material surfaces, such as soft textiles, elastic polymers, or temperature-responsive textures, can enable the system to simulate a broader emotional palette through differentiated squeeze-based haptic sensations~\cite{han2024ambient}.

By coupling real-time physiological sensing with these squeeze-based haptic innovations, future systems could deliver emotionally intelligent and materially expressive haptic experiences. These developments would not only deepen emotional resonance but also pave the way for emotionally supportive VR environments that are responsive, adaptive, and sustainable across diverse use contexts and populations, from relaxation and coping in the general population~\cite{riches2021virtual,liszio2018relaxing} to regulation in challenging contexts and populations~\cite{10.1145/3613905.3637139}.


\subsection{Chronic Stress in the Post-epidemic Era and the Application Value of VR-Integrated Squeeze Interaction Paradigm}
Chronic stress has transcended individual health issues and become a major public health challenge affecting global socioeconomic development. Especially in the post-pandemic era, long-term concerns about public health security, coupled with emerging change, such as weakened interpersonal connections and blurred boundaries between work and life that are triggered by past experience of remote work and social isolation, have not only made the inducing factors of chronic stress more complex but also led to a wider penetration of its impact among the population \cite{feng2025relationship}. The complexity of chronic stress necessitates the establishment of a multi-level prevention and intervention system. At the policy level, it is necessary to improve the mental health service network, while enterprises should proactively provide psychological support for employees and individuals need to master scientific and effective stress management skills \cite{kopp2007chronic}. Only through systematic and collaborative responses can the spread of this ``invisible epidemic'' be effectively curbed.

The core challenge in alleviating chronic stress lies in breaking through the limitation of the non-sustainable effects of short-term emotional comfort \cite{mcewen2017neurobiological}. The establishing of controllable regulation mechanism is particularly critical for individuals with chronic stress \cite{kant1992effects, worley2018prefrontal}, and we observe that this demand is partially reflected in the design logic of squeeze interactions.

On the one hand, as noted earlier, this system demonstrates the potential to deliver dynamically responsive, growth-oriented emotional interventions by coupling differentiated squeeze-based haptic with real-time physiological sensing. Given the persistent nature of many stressors~\cite{schetter2011resilience}, effective chronic stress management needs to be staged and developmental in phases~\cite{bhagya2017short}. Building on VR Calm Plus, future systems can further refine these squeeze-based haptic and sensing channels to move beyond traditional interventions with fixed scenarios and repetitive experiences, enabling long-term, phase-specific support for chronic stress.

On the other hand, it helps users reestablish a certain sense of control over their physical and emotional states. Our qualitative research findings indicate that users can gain a strong sense of control during squeezing actions aligned with embodied cognition, which may be attributed to the aforementioned ``emotional amplification'' effect and improved interaction transparency. Further refining and leveraging the system's sensory coherence and interaction transparency design in the future will enable users to clearly perceive the causal link between their behaviors and emotions. Additionally, rhythmic and tangible squeezing interaction training accompanied by sensory feedback (similar to the stage 2) may help address anxiety-related somatization symptoms induced by long-term chronic stress. Taking the squeezable plush device in the VR Calm Plus as an example, squeezing the device can engage muscles prone to somatic anxiety, such as forearm flexors, biceps brachii, and deltoid muscles. The core dilemma in people's inability to autonomously alleviate somatic anxiety lies in being trapped in a vicious cycle of ``emotional anxiety → somatic symptoms → excessive focus on somatic symptoms → increased anxiety'' \cite{mallorqui2016mind}, resulting in the loss or partial loss of control over their physical and emotional states. In this context, further refining the tangible input interaction mechanism in the future may break the physiological chain reaction of ``anxiety → muscle tension'' by establishing new causal links between actions and emotions. Furthermore, research findings indicate that participants' attention to the interaction experience itself was significantly enhanced, which may weaken the cycle of ``focus on somatic symptoms → increased anxiety'' by shifting attention from internal somatic symptoms to external interaction experiences \cite{sahm2024putting}.

In conclusion, the VR-integrated squeeze interaction model represented by VR Calm Plus has demonstrated potential advantages for further exploration in both professional and non-professional settings. Based on the current cross-sectional experimental data and self-report results, it may develop into an accessible and reliable tool for short-term stress regulation among specific populations. At present, it only provides a preliminary exploratory solution paradigm for addressing stress-related challenges in specific contexts. However, future research must supplement long-term follow-up data, expand to more diverse samples, and verify its effectiveness in real-world settings.

\section{Limitations, Challenges and Future Work}
While this study offers promising evidence for the role of squeeze feedback in enhancing emotion regulation in VR, several limitations warrant consideration. First, all participants were from China. Although there is no conclusive evidence suggesting significant physiological differences in squeeze perception across ethnic or cultural groups, the demographic homogeneity may limit the generalizability of the findings. In particular, emotional responses to immersive environments and squeeze feedback may vary across cultures due to differing norms around affect display, regulation strategies, and interpretations of multisensory input~\cite{esposito2015needs}. For example, cultures with more restrained emotional expression may report lower self-assessed positive affect even in similarly designed interventions, while collectivist versus individualist values may shape regulation preferences and the perceived acceptability of tactile interaction~\cite{markus2014culture}. Future research should include more diverse populations to examine the cross-cultural applicability of squeeze-based emotion regulation. 
Participants reported being able to operate the device subconsciously (P5), and all indicated a complete understanding of the operational mechanism of the squeeze interaction. This indicates that compared with other complex interaction mechanisms, this embodied cognition-integrated interaction method may be more suitable for users with limited prior experience in similar interventions, but participants in this study were concentrated in young and middle-aged groups, with only a minimal number of elderly participants (P2 Age=66). The adaptability to the elderly population (who may have potential operational barriers) and child population (who may exhibit distinct attention spans and interaction preferences) remains to be verified in future work by expanding the sample size to confirm whether the advantages can be sustained. Second, the experiment was conducted in a controlled laboratory setting, which may not fully capture how users interact with the system in everyday contexts. With repeated exposure in real-world environments, users may experience habituation, whereby the emotional and physiological effects of the intervention diminish over time~\cite{kyriakou2019detecting}. Evaluating the system in real-world environments, such as home use or clinical applications, could provide further insights into its ecological validity and long-term effectiveness. Finally, from a practical perspective, the long-term durability of the plush interface remains a challenge. While no damage was observed in this study, extended use may reduce the toy's softness and squeeze responsiveness. Future designs should include stress testing and collect user feedback to ensure lasting comfort and performance.

\section{Conclusion}
This study presents the design and evaluation of VR Calm Plus, a VR system enhanced with squeeze interaction for emotion regulation. Through a combination of validated emotion scales (PANAS-X), a customized subjective experience questionnaire, and semi-structured interviews, we quantitatively and qualitatively assessed the emotional and experiential impacts of squeeze feedback in immersive environments. Our findings reveal that integrating squeeze interaction, specifically, a squeezing-based tangible interface, significantly enhances users' emotional engagement, attentional focus, and calming affective states. Participants in the Squeeze Interaction condition not only reported improved serenity and attentiveness, but also experienced greater interaction clarity, emotional resonance, and a heightened sense of control.

These results underscore the role of embodied interaction in affective VR design, and support the use of squeeze feedback as both an emotional amplifier and a mechanism for enhancing interaction transparency. Drawing from theories of embodied cognition and media richness, our findings highlight the importance of multisensory coherence in facilitating emotionally intelligible and user-friendly experiences. Moreover, the proposed squeezable interface demonstrates how even simple physical gestures, when aligned with audiovisual feedback, can reinforce emotional understanding and promote active regulation.

Looking ahead, our work opens several avenues for future research. We advocate for the development of bi-directional emotion-squeezable systems, where squeeze feedback is dynamically modulated by users' physiological signals, enabling closed-loop emotion support. We hope this work could contributes to a growing body of research exploring how affective computing and multisensory interaction can converge to support well-being and self-regulation in virtual environments.


\begin{acks}
The authors wish to thank our participants and anonymous reviewers. This work was funded by an anonymous sponsor and approved by the IRB of an anonymous institution. To enhance transparency and uphold ethical standards, the authors disclose the following use of generative AI tools: we used large language models (e.g., ChatGPT) only to check and improve grammar and spelling in the text; all experimental results were obtained from real human participants. This work was supported by the Beijing Natural Science Foundation-Youth Project (Grant No. 4254082).
\end{acks}

\balance
\bibliographystyle{ACM-Reference-Format}
\bibliography{references, main}

\appendix
\section*{Appendix}
\begin{table*}[htbp]
\centering
\begingroup
\small
\renewcommand{\arraystretch}{1.0} 
\setlength{\tabcolsep}{15pt}       
\begin{tabularx}{\linewidth}{c c c L{0.25\linewidth} Y}
\toprule
\textbf{Participant No.} & \textbf{Age} & \textbf{Gender} & \textbf{Meditation Experience} & \textbf{VR Experience} \\
\midrule
P1  & 28 & Male   & Basic Understanding & Initial Exposure \\
\rowcolor[HTML]{EFEFEF}
P2  & 66 & Female & Never               & Never \\
P3  & 48 & Female & Never               & Never \\
\rowcolor[HTML]{EFEFEF}
P4  & 26 & Female & Basic Understanding & Basic Understanding \\
P5  & 22 & Female & Initial Exposure     & Initial Exposure \\
\rowcolor[HTML]{EFEFEF}
P6  & 49 & Female & Never               & Basic Understanding \\
P7  & 19 & Female & Proficient          & Never \\
\rowcolor[HTML]{EFEFEF}
P8  & 24 & Female & Initial Exposure     & Proficient \\
P9  & 24 & Male   & Initial Exposure     & Proficient \\
\rowcolor[HTML]{EFEFEF}
P10 & 26 & Male   & Initial Exposure     & Initial Exposure \\
P11 & 24 & Female & Initial Exposure     & Basic Understanding \\
\rowcolor[HTML]{EFEFEF}
P12 & 25 & Female & Basic Understanding & Never \\
P13 & 22 & Male   & Initial Exposure     & Basic Understanding \\
\rowcolor[HTML]{EFEFEF}
P14 & 55 & Female & Never               & Basic Understanding \\
P15 & 24 & Male   & Never               & Initial Exposure \\
\rowcolor[HTML]{EFEFEF}
P16 & 21 & Male   & Basic Understanding & Initial Exposure \\
P17 & 23 & Male   & Never               & Initial Exposure \\
\rowcolor[HTML]{EFEFEF}
P18 & 23 & Female & Never               & Basic Understanding \\
P19 & 24 & Female & Initial Exposure     & Initial Exposure \\
\rowcolor[HTML]{EFEFEF}
P20 & 25 & Male   & Never               & Basic Understanding \\
P21 & 21 & Male   & Never               & Basic Understanding \\
\rowcolor[HTML]{EFEFEF}
P22 & 47 & Female & Never               & Never \\
P23 & 20 & Male   & Never               & Never \\
\rowcolor[HTML]{EFEFEF}
P24 & 31 & Male   & Experienced         & Proficient \\
P25 & 20 & Female & Initial Exposure     & Never \\
\rowcolor[HTML]{EFEFEF}
P26 & 21 & Female & Never               & Never \\
P27 & 19 & Female & Initial Exposure     & Initial Exposure \\
\rowcolor[HTML]{EFEFEF}
P28 & 22 & Male   & Basic Understanding & Basic Understanding \\
P29 & 26 & Female & Basic Understanding & Initial Exposure \\
\rowcolor[HTML]{EFEFEF}
P30 & 26 & Male   & Never               & Initial Exposure \\
P31 & 33 & Male   & Proficient          & Basic Understanding \\
\rowcolor[HTML]{EFEFEF}
P32 & 39 & Female & Experienced         & Basic Understanding \\
P33 & 20 & Female & Never               & Initial Exposure \\
\rowcolor[HTML]{EFEFEF}
P34 & 23 & Female & Basic Understanding & Basic Understanding \\
P35 & 24 & Male   & Initial Exposure     & Initial Exposure \\
\rowcolor[HTML]{EFEFEF}
P36 & 36 & Female & Proficient          & Basic Understanding \\
P37 & 25 & Female & Proficient          & Proficient \\
\rowcolor[HTML]{EFEFEF}
P38 & 33 & Male   & Basic Understanding & Initial Exposure \\
P39 & 22 & Male   & Experienced         & Initial Exposure \\
\rowcolor[HTML]{EFEFEF}
P40 & 27 & Male   & Initial Exposure     & Initial Exposure \\
\bottomrule
\multicolumn{5}{@{}l}{\textbf{Experience Categories:}} \\
\multicolumn{5}{@{}l}{\textit{Never}: No prior exposure or practice.} \\
\multicolumn{5}{@{}l}{\textit{Initial Exposure}: One-time or very limited exposure.} \\
\multicolumn{5}{@{}l}{\textit{Basic Understanding}: Some conceptual knowledge but limited practice.} \\
\multicolumn{5}{@{}l}{\textit{Experienced}: Regular practice or multiple applications.} \\
\multicolumn{5}{@{}l}{\textit{Proficient}: Advanced practice or expertise through long-term use.} \\
\end{tabularx}
\endgroup
\Description{The demographic characteristics show the diversity of VR and meditation experience among participants, which indicate that 40\% of them are novices to both VR and meditation.}
\caption{Demographic characteristics of study participants.}
\label{table:demograph_participants}
\end{table*}

\end{document}